\documentstyle[12pt,epsf,amssymb]{article}
\catcode`\@=11
\def \thesection {\arabic{section}.}

\def \be  {\begin{equation}}
\def \ee  {\end{equation}}
\def \ba  {\begin{eqnarray}}
\def \ea  {\end{eqnarray}}
\def \baa {\begin{eqnarray*}}
\def \eaa {\end{eqnarray*}}
\def \bb  {\begin{thebibliography}}
\def \eb  {\end{thebibliography}}
\def \lab #1 {\label{#1}}
\newcommand \ci [1] {\cite{#1}}
\newcommand \bi [1] {\bibitem{#1}}
\newcommand\re[1]{(\ref{#1})}
\def \qqquad {\qquad\quad}

\def \matrix #1 {\left(\begin{array}{cc} #1 \end{array}\right)}

\def \Im {\mathop{\rm Im}\nolimits}
\def \Re {\mathop{\rm Re}\nolimits}

\def \e  {\mathop{\rm e}\nolimits}

\newcommand\lr[1]{{\left({#1}\right)}}
\newcommand \widebar [1] {\overline{#1}}

\newcommand \VEV [1] {\left\langle{#1}\right\rangle}
\newcommand \ket [1] {|{#1}\rangle}

\newcommand{\as}{\ifmmode\alpha_{\rm s}\else{$\alpha_{\rm s}$}\fi}

\def \CO {{\cal O}}

\font\cmss=cmss12 
\def\inbar{\,\vrule height1.5ex width.4pt depth0pt}
\def\IC{\relax\hbox{$\inbar\kern-.3em{\rm C}$}}
\def\IZ{\relax{\hbox{\cmss Z\kern-.4em Z}}}
\def\IR{{\hbox{{\rm I}\kern-.2em\hbox{\rm R}}}}
\def\IP{{\hbox{{\rm I}\kern-.2em\hbox{\rm P}}}}
\def\II{\hbox{{1}\kern-.25em\hbox{l}}}
\def\Ir{{\tiny \IR}}
\def\Iz{{\tiny \hbox{\rm Z}}}
\def\IK{{\hbox{{\rm I}\kern-.2em\hbox{\rm K}}}}

\def\numberbysection{\@addtoreset{equation}{section}
                     \def\theequation{\thesection\arabic{equation}}}
\numberbysection

\begin{document}

\def\thefootnote{\fnsymbol{footnote}}
\thispagestyle{empty}

\hfill\parbox{50mm}{{\sc LPTHE--Orsay--97--73} \par
                         hep-ph/9801377       \par
                         December, 1997}
\vspace*{35mm}
\begin{center}
{\LARGE WKB quantization of reggeon compound states\\[5mm] in high-energy QCD}
\par\vspace*{15mm}\par
{\large G.~P.~Korchemsky} 
\par\bigskip\par\medskip
 
{\em Laboratoire de Physique Th\'eorique et Hautes Energies%
\footnote{Laboratoire associ\'e au Centre National de la Recherche
Scientifique (URA D063)} \\
Universit\'e de Paris XI, Centre d'Orsay, b\^at. 211\\
91405 Orsay C\'edex, France}
\par\medskip
{\em Laboratory of Theoretical Physics,\\
Joint Institute for Nuclear Research, \\
141980 Dubna, Russia}
\end{center}

\vspace*{20mm}

\begin{abstract}
We study the spectrum of the reggeized gluon compound states in high-energy 
QCD. The energy of the states depends on the total set of quantum numbers 
whose values are constrained by the quantization conditions. To establish 
their explicit form we apply the methods of nonlinear WKB analysis to the 
Schr\"odinger equation for the $N$ reggeon state in the separated coordinates. 
We solve the quantization conditions for the $N=3$ reggeon states in the
leading order of the WKB expansion and observe a good agreement of 
the obtained spectrum of quantum numbers with available exact solutions.
\end{abstract}

\newpage

\def\thefootnote{\arabic{footnote}}
\setcounter{footnote} 0

\section{Introduction}
In perturbative QCD analysis of the Regge asymptotics of
hadronic scattering amplitudes, the power rise with energy
of the physical cross-sections is attributed to the contribution
of the color-singlet compound gluonic states. These states
are built from an arbitrary conserved number, $N=2,3,...$, of reggeized
gluons, or reggeons, and they can be defined as solutions of the effective
$(2+1)-$dimensional Schr\"odinger equation \ci{BKP}
\be
{\cal H}_N \ket{\chi_{N,\{q\}}} = E_{N,\{q\}} \ket{\chi_{N,\{q\}}}\,.
\lab{BKP}
\ee
The effective QCD Hamiltonian ${\cal H}_N$ describes a pair-wise 
interaction between $N$ reggeons on the 2-dimensional plane
of their transverse coordinates, $\vec\rho_k=(\rho_{x_k},\rho_{y_k})$
with $k=1$, $...$, $N$, and its explicit form can be found in \ci{Lip1}.
The longitudinal gluonic degrees of freedom define the rapidity of 
reggeons that enter into \re{BKP} as a evolution time. The wave 
function $\chi_{N,\{q\}}(\vec\rho_1-\vec\rho_0,...,\vec\rho_N-\vec\rho_0)$
depends on some set of quantum numbers $\{q\}$ 
and the center-of-mass coordinate $\vec\rho_0$ of the reggeon
compound state. 

The contribution of the $N$ reggeon states to the
physical cross section has a form 
$$
\sigma (s) \sim  s^{\frac{\as N_c}{2\pi}E_{N,\{q\}}}
$$ 
and at large energies $s$ it is dominated by the state with the
maximal energy, $E_0=\max_{\{q\}} E_{N,\{q\}}$. Therefore
solving the Schr\"odinger equation \re{BKP} one is mostly interesting in
finding the state with the energy $E_0$, or equivalently the
ground state of the hamiltonian $-{\cal H}_N$.

The Schr\"odinger equation \re{BKP} has a number of remarkable properties.
The effective QCD Hamiltonian ${\cal H}_N$ is invariant under projective 
$SL(2,\IC)$ transformations of the 2-dimensional gluon coordinates 
$\rho_j$ \ci{Lip1}
\be
z_j=\rho_{x_j}+i\rho_{y_j}\,,\qquad
\bar z_j=z_j^*\,,\qquad
z_j\to \frac{az_j+b}{cz_j+d}
\lab{SL2}
\ee
with $ad-bc=1$ and $(z_j,\bar z_j)$ being (anti)holomorphic reggeon
coordinates. Denoting the generators of these transformations as
\be
L_0=\sum_{k=1}^N z_k   \partial_k\,,\qquad
L_-=\sum_{k=1}^N       \partial_k\,,\qquad
L_+=\sum_{k=1}^N z_k^2 \partial_k\,,\qquad
\lab{L's}
\ee
and having similar expressions for the antiholomorphic generators, one
may verify that $[{\cal H}_N,L_\alpha]=[{\cal H}_N,\bar L_\alpha]=0$.
This symmetry allows to classify possible solutions to \re{BKP}
according to the representations of the $SL(2,\IC)$ group. In
particular, the wave function is transformed under projective
transformations as
\be
\chi_{N,\{q\}} \to (cz_0+d)^{2h} (\bar c\bar z_0+\bar d)^{2\bar h}
\chi_{N,\{q\}}
\lab{pro}
\ee
with $(z_0,\bar z_0)$ being (anti)holomorphic coordinates of the
center-of-mass. The conformal weights of the state, $h$ and $\bar h$,
take the values corresponding to the unitary principal series
representation of the $SL(2,\IC)$ group \ci{group}
\be
h=\frac{1+n}2+i\nu\,,\qquad
\bar h=\frac{1-n}2+i\nu\,,\qquad
(n=\IZ\,,\quad \nu=\IR)
\lab{h}
\ee
parameterized by the pair $(n,\nu)$ of integer $n$ and
real $\nu$. These parameters determine the spin $s=n$ and
the scaling dimension $d=1+2i\nu$ of the state and control the
transformation properties of the state under rotations,
$z_k\to \e^{i\phi} z_k$, $\bar z_k\to \e^{-i\phi} \bar z_k$,
$\chi_N\to\e^{in\phi}\chi_N$ and dilatations, $z_k\to\lambda z_k$,
$\bar z_k\to\lambda\bar z_k$, $\chi_N\to\lambda^{-1-2i\nu}\chi_N$.

Being expressed in terms of holomorphic and
antiholomorphic reggeon coordinates, the effective Hamiltonian
splits into the sum of two mutually commuting $(1+1)-$dimensional
Hamiltonians acting separately on (anti)holomorphic coordinates,
${\cal H}_N=H_N+\widebar H_N$ and $[H_N,\widebar H_N]=0$ \ci{Lip2}.
This property is exact for $N=2$ and $N=3$ reggeon states
and for $N\ge 4$ it holds in the multi-color limit of QCD. It
implies that the original equation \re{BKP} can be replaced
by the system of two $(1+1)-$dimensional Schr\"odinger equations
in holomorphic and antiholomorphic subspaces. As a result, the wave
function of the state can be factorized as
\be
\chi_N(z,\bar z)=\sum_{\alpha} C_{\alpha}\,
\varphi_{N,\alpha}(z) \bar\varphi_{N,\alpha}(\bar z)\,,
\lab{wf}
\ee
where ${z}=(z_1-z_0,...,z_N-z_0)$ denotes the set of $N$ holomorphic
coordinates and the coefficients $C_\alpha$ are fixed by the condition for
$\chi_N(z,\bar z)$ to be a well defined function of its arguments on the
2-dimensional plane of transverse coordinates. Additionally, the energy $E_N$
of the state can be evaluated as the sum of holomorphic and antiholomorphic
energies that one finds solving two $(1+1)-$dimensional Schr\"odinger 
equations. 

It turns out that the holomorphic Hamiltonian $H_N$ coincides with the 
Hamiltonian of 1-dimensional Heisenberg magnet of noncompact $SL(2,\IC)$ 
spin zero \ci{FK,Lip}. There is however an important difference 
between the holomorphic Hamiltonian $H_N$ and well-known integrable 
$(1+1)-$dimensional Hamiltonians \ci{SoV,KK}. Examining the properties of 
$H_N$ one finds \ci{Bet} that in contrast with the full $(2+1)-$dimensional 
effective Hamiltonian, the operator $H_N$ is unbounded on the space of the wave 
functions \re{pro}. The same property holds for the antiholomorphic 
hamiltonian $\widebar H_N$. It is only the sum of holomorphic and 
antiholomorphic $(1+1)-$dimensional Hamiltonians that is bounded from 
above. This happens because the holomorphic and antiholomorphic energies 
depend on the common set of quantum numbers \re{h} and infinities cancel 
in their sum.

Analyzing the Schr\"odinger equation \re{BKP} one finds \ci{FK,Lip} that it 
possesses the family of mutually commuting holomorphic and antiholomorphic 
integrals of motion, $\{q_k,\bar q_k\}$,
$k=2,...,N$ defined as
\be
q_k=i^k\sum_{1\le j_1 < ...<j_k \le N} z_{j_1j_2}...z_{j_kj_1}
\partial_{j_1}... \partial_{j_k} \,,\qquad
\qquad [q_k,{\cal H}_N]=0\,,
\lab{q}
\ee
where $z_{jk}=z_j-z_k$ and the definition of the antiholomorphic charges
$\bar q_k$ is similar with $z_j$ replaced by $\bar z_j$. The total number
of the conserved charges, $2(N-1)$, is large enough for the $N-$gluon
Schr\"odinger equation \re{BKP} to be completely integrable. The wave
function of the $N-$reggeon state, $\chi_{N,\{q\}}$, diagonalizes these 
charges and one identifies the corresponding eigenvalues as a complete set
of quantum numbers $\{q\}$ introduced before in \re{BKP}. One also
takes into account that on the space of the wave functions \re{pro} 
the holomorphic and antiholomorphic charges are related as \ci{Bet}
$$
\bar q_k = q_k^*\,,\qquad k=2,...,N\,.
$$

The integrability of the Schr\"odinger equation \re{BKP} ensures that the
effective Hamiltonian ${\cal H}_N$ acting on the $2N$ coordinates
of reggeons can be expressed as a complicated function of $2(N-1)$
integrals of motion $(q_k,\bar q_k)$ plus $2$ operators reserved for the
components of the center-of-mass momentum of the state. We notice
however that (anti)holomorphic components of the total momentum coincide
with the generators, $L_-$ and $\bar L_-$, of the projective
transformations \re{L's} and the dependence of the Hamiltonian on them
is protected by the $SL(2,\IC)$ symmetry, $[{\cal H}_N, L_\alpha]=0$.
This leads to the following identity \ci{Bet}
$$
E_{N,\{q\}}= \varepsilon_N(q_2, q_3, ..., q_N)
            +\varepsilon_N(\bar q_2,\bar q_3, ...,\bar q_N)
	    \,,
$$
where $\varepsilon_N$ denotes the energy of the (anti)holomorphic Hamiltonian.

Thus, the problem of calculating the spectrum of the effective QCD Hamiltonian
\re{BKP} is reduced, first, to finding the explicit form of $\varepsilon_N$ 
as a real function of the conserved charges $q_2$, $...$, $q_N$ and, second, to
establishing the quantization conditions for their eigenvalues. The first
part of the problem was studied using different approaches \ci{Qua,WJ} and
similar results for $\varepsilon_N$ were obtained. In the present
paper we address the problem of finding the spectrum of the conserved
charges. 

The paper is organized as follows. In the Sect.~2 we employ the WKB
expansion to obtain the quantization conditions for the eigenvalues of
conserved charges. We show in Sect.~3 that these conditions have 
the form of the Whitham equations. Solving the Whitham equations in
the special case of $N=3$ reggeon states we obtain the quantized 
charges $q_2$ and $q_3$ and study their properties in Sect.~4. Concluding
remarks can be found in Sect.~5. Appendix contains some properties of 
the elliptic integrals which are helpful in solving the Whitham equation.

\section{Quantization conditions}

The ``lowest'' integrals of motion, $q_2$ and $\bar q_2$, provide the
simplest example of the quantization conditions. One verifies using
\re{L's} and \re{q} that $q_2$ and $\bar q_2$ coincide with the quadratic
Casimir operators of the $SL(2,\IC)$ group, $q_2=L_0(L_0+1)+L_-L_+$, and for
the wave functions transforming as \re{pro} their eigenvalues take the form
\be
q_2=-h(h-1)=\frac14-\lr{\frac{n}2+i\nu}^2\,,\qquad
\bar q_2=-\bar h(\bar h-1)=q_2^*\,,
\lab{q2}
\ee
being discrete in $n$ and continuous in $\nu$. The quantization conditions
for ``higher'' integrals of motion $q_k$ and $\bar q_k$ ($k\ge 3$) are
more involved \ci{Bet,Qua}. One can get some insight into their properties
by exploring the relation between the wave functions of reggeon
compound states $\chi_N(z,\bar z)$ and 2-dimensional conformal field theories.
To this end, let us consider the following function \ci{FK,Qua}
$$
\widehat\chi_N(z,\bar z)=(z_{12}z_{23}...z_{N1})^{-1} \chi_N(z,\bar z)\,.
$$
According to its properties under the projective transformations, \re{pro}, the
function $\widehat\chi_N$ can be represented as $(N+1)-$point correlation
function in 2-dimensional conformal field theory
\be
\widehat\chi_N(z,\bar z)=
\VEV{\phi(z_1,\bar z_1)...\phi(z_N,\bar z_N) O_{h,\bar h}(z_0,\bar z_0)}\,.
\lab{cf}
\ee
Here, $\phi(z,\bar z)$ is a quasiprimary reggeon field with the conformal
weights $(1,0)$ and $O_{h,\bar h}(z_0,\bar z_0)$ is a quasiprimary operator
interpolating the $N$ reggeon compound state. The operator $O_{h,\bar h}$
has the conformal weights \re{h} and it is
transformed under \re{SL2} according to the irreducible principal series
representation (irreps) of the $SL(2,\IC)$ group labeled as $t^{(\nu,2n)}$.
For the reggeon operator $\phi(z,\bar z)$ the corresponding irreps can be
identified as $t^{(0,2)}$. The product of $N$ reggeon operators in \re{cf}
is reducible with respect to the action of the $SL(2,\IC)$ transformations.
It corresponds to the tensor product of $N$ copies of $t^{(0,2)}$
which in turn can be decomposed into irreducible components using the
following fusion rules \ci{group}
$$
t^{(0,2)}\otimes t^{(0,2)}=\bigoplus_{n\in\Iz\,,\nu\in\Ir} t^{(\nu,2n)}
\,,\qquad
t^{(\nu,2n)}\otimes t^{(0,2)}=\bigoplus_{m\in\Iz\,,\rho\in\Ir} t^{(\rho,2m)}\,.
$$
In conformal field theory these relations are implemented in the operator
product expansion of the quasiprimary operators $\phi(z,\bar z)$ and
$O_{h,\bar h}(z,\bar z)$ \ci{Tri,SL2}. Applying the fusion rules $N-1$ times 
one obtains an expansion of the product  
$\phi(z_1,\bar z_1)...\phi(z_N,\bar z_N)$ in \re{cf}
into an infinite sum of the irreducible components. Each component can be
parameterized in two equivalent ways: either by the conformal weights 
$(h,\bar h)$ plus the set of $2(N-2)$ quantum numbers $(q_k,\bar q_k)$, or 
by the set of $2(N-1)$ numbers $(n_k,\nu_k)$ corresponding to the $N-1$ 
intermediate states $t^{(\nu_k,2n_k)}$. This leads to the following dependence
\be
 q_k = q_k\lr{(n_1,\nu_1),...,(n_{N-1},\nu_{N-1})}\,,
\qquad
(k=2,...,N)
\lab{Q-gen}
\ee
together with $h=\frac{1+n_{N-1}}2+i\nu_{N-1}$ and $\bar h=1-h^*$.
Thus, apart from the set $(n_{N-1},\nu_{N-1})$ defining the conformal
weight of the $N$ reggeon state, one expects the appearance of $(N-2)$
additional sets $(n_k,\nu_k)$ parameterizing the eigenvalues of the
operators $q_k$ and $\bar q_k$. In particular, for given conformal weights
$(h,\bar h)$ the possible values of the quantum numbers $q_k$ and $\bar q_k$
form the family of continuous functions of real $\nu_1$, $...$, $\nu_{N-2}$,
different members of which are labeled by integers $n_1$, $...$, $n_{N-2}$.

Let us consider two special cases. In the first case, one takes all $\nu_k=0$
and $n_k={\rm odd}$. Using \re{h} one finds that it corresponds to restricting
the conformal weights of the intermediate states $t^{(\nu_k,2n_k)}$ and
the $N$ reggeon state, $t^{(\nu_{N-1},2n_{N-1})}$, to take only integer values. 
The spectrum of the
operators $q_k$ and $\bar q_k$ is expected to be discrete. Indeed, this
result was previously obtained within the algebraic Bethe Ansatz approach
\ci{FK} by numerical solution of the Bethe equations \ci{Bet}
and later the analytical expressions for quantized $q_k$ and $\bar q_k$ were
derived in the quasiclassical approximation \ci{Qua}.
In the second case, one chooses $n_k=0$ and
$\nu_k$ are arbitrary. The conformal weights of the intermediate states
have the form $h_k=\frac12+i\nu_k$ and they run along imaginary axis.
It has been argued \ci{Bet} that it is for these values of the conformal
weights that the energy of the $N$ reggeon states approaches its maximal value.
The quantized $q_k$ and $\bar q_k$ become continuous functions of real
$\nu_1$, $...$, $\nu_{N-1}$. As was shown in \ci{Int}, these functions
can be found within the WKB approach as solutions of the Whitham equations. 
In the next Section we derive these equations, solve them 
and study the properties of quantized $q_k$ and $\bar q_k$.

\subsection{Separated variables}

The eigenvalue problem for the operators $q_k$ and $\bar q_k$ defined in \re{q}
leads to a complicated system of partial differential equations of the
$N-$th order on the wave function $\chi_N$. One can significantly simplify the
problem by using the fact that quantization of the charges $q_k$ and
$\bar q_k$ does not depend on the particular representation in which the
wave function $\chi_N$ is defined in order to go from $(z,\bar z)-$coordinates
to a new set of {\it separated\/} holomorphic reggeon coordinates \ci{SoV,KK},
$x_1$,...$x_N$, and their antiholomorphic counterparts $\bar x_k=x_k^*$.

In the separated coordinates the wave function of the $N$ reggeon state
is factorized into the product of functions each depending on only one
coordinate
\be
\chi_N(x,\bar x)=Q(x_1)...Q(x_{N-1})\, x_N^{-h} \times
\widebar Q(\bar x_1)...\widebar Q(\bar x_{N-1})\,
\bar x_N^{-\bar h}\,.
\lab{chi}
\ee
Here, the function $Q(x)$ is a solution of the second order finite-difference
equation \ci{FK}
\be
Q(x+i)+Q(x-i)=x^{-N}\Lambda(x)  Q(x)\,,\qquad
\Lambda(x)=2x^N+q_2x^{N-2}+...+q_N
\lab{Beq}
\ee
known as the Baxter equation and $\widebar Q(x)$ satisfies similar relation
with all $q_k$ replaced by $\bar q_k=q_k^*$.

The separated gluon coordinates have the following properties. The
coordinates $x_N$ and $\bar x_N$ correspond to the center-of-mass motion
and they are defined as eigenvalues of the operators
\be
x_N=L_0L_-^{-1} \,,\qquad
\bar x_N= \bar L_0 \bar L_-^{-1}
\lab{xN}
\ee
with $L_-$ and $L_0$ being the $SL(2,\IC)$ generators \re{L's}. The definition
of the remaining coordinates $x_k$ and $\bar x_k$ for $k=1,..,N-1$ is more
involved and it can be found in \ci{Bet,SoV}. It will be important for us that they
satisfy the relations \ci{SoV}
$$
[L_0,x_k]=[L_-,x_k]=0\,,\qquad
[L_0,x_N]=ix_N\,,\qquad [L_-,x_N]=-i
$$
with $k=1$, $...$, $N-1$ and similar relations hold for $\bar x_k$ and $\bar 
x_N$. From these relations one finds that the separated coordinates $x_k$ 
and $\bar x_k$, being functions of the gluon $(z_j,\bar z_j)-$coordinates,  
are invariant under translations, $z_k\to z_k+\epsilon$, and dilatations, 
$z_k\to \lambda z_k$, while the coordinates $x_N$ and $\bar x_N$ are transformed 
in the same way as $z_k$ and $\bar z_k$. Therefore, in the expression \re{chi} 
for the wave function in separated coordinates, it is the factor
$x_N^{-h}\bar x_N^{-\bar h}$ that carries the spin $n=h-\bar h$ and the scaling
dimension $d=1+2i\nu=h+\bar h$ of the $N$ reggeon state.
Moreover, making the shift, $z_k\to z_k-z_0$, one can restore the
dependence of the wave function \re{chi} on the center-of-mass coordinate
of the state, $(z_0,\bar z_0)$, as $x_N\to x_N-z_0$ and $x_k\to x_k$
$(k=1,...,N-1)$.

The wave function \re{wf} has to be a well defined function of the reggeon
coordinates $z_k$ and $\bar z_k$. Performing transition to the separated
coordinates one imposes the same condition on \re{chi} as a function
of the complex $x_k$ and $\bar x_k=x_k^*$. Let us first examine the
dependence of $\chi_N$ on the coordinate $x_N$. If $\chi_N$ is a single
valued function of complex $x_N$ it should not be changed as 
$x_N\to\e^{2\pi i}x_N$ in \re{chi}. One checks that this is true provided 
that the spin $h-\bar h=n$ of the state is integer. Requiring additionally
that the scaling dimension of the state has to be $d=h+\bar h=1+2i\nu$ 
one reproduces the quantization conditions \re{h} for the conformal weights.
We notice that although the wave function \re{chi} stays invariant under 
$2\pi$ rotations on the $x_N-$plane, its holomorphic and antiholomorphic 
components get nontrivial monodromies for all values of the conformal weights 
$h$ and $\bar h$ except for integer ones. The latter case corresponds to the 
polynomial solutions studied in \ci{Bet,Qua}. 

To satisfy the condition of singlevaluedness for the dependence
of \re{chi} on the remaining coordinates $x_k$ one has to examine the
properties of the product $Q(x)\widebar Q(x^*)$ on the complex $x-$plane. 
As we will show below, this consideration leads to the 
selection rules for the quantum numbers $q_3$, $...$, $q_N$ 

\subsection{Properties of the Baxter equation}

The function $Q(x)$ entering \re{chi} obeys the Baxter equation \re{Beq}. 
This equation has two linear independent solutions, $Q_+(x)$ and $Q_-(x)$, 
satisfying the normalization condition
$$
W(x)=Q_+(x+i)Q_-(x)-Q_-(x+i)Q_+(x)=1
$$
and having the following asymptotics at infinity
\be
Q_+(x) \sim x^h \,,\qquad Q_-(x) \sim x^{1-h} \qquad
\mbox{as $x\to\infty$}\,.
\lab{as}
\ee
The analysis of the Baxter equation for the function $\widebar Q(\bar x)$
is similar since the functions $(Q(x))^*$ and $\widebar Q(x^*)$ satisfy the 
same equation. As a result, one gets the following general expression
$$
Q(x) \widebar Q(x^*)=\sum_{\alpha,\beta=\pm} C_{\alpha\beta}\, Q_\alpha(x)
(Q_\beta(x))^*
$$
with $C_{\alpha\beta}$ being arbitrary coefficients. However, it follows 
from \re{as} that for the conformal weights $h$ and $\bar h$ of the form 
\re{h} the functions $Q_+(x)$ and $Q_-(x)$ acquire nontrivial monodromies 
at infinity as $x\to \e^{2\pi i} x$, which should cancel in the product
$Q(x)\widebar Q(x^*)$. This requirement leads to 
$C_{--}=C_{++}=0$ and 
\be
Q(x) \widebar Q(x^*)=C_{+-} Q_+(x) (Q_-(x))^* + C_{-+} Q_-(x) (Q_+(x))^*\,.
\lab{ans}
\ee
with $C_{+-}$ and $C_{-+}$ being some constants.
 
The asymptotic behaviour \re{as} of the functions $Q_\pm(x)$
is fixed by the conformal weight of the state and it is not
sensitive to the value of ``higher'' quantum numbers $q_3$, $...$, $q_N$.
To obtain the quantization conditions for $q_3$, $...$, $q_N$ one has
to identify the remaining singularities of the solutions of the Baxter
equation, $Q_+(x)$ and $Q_-(x)$, on the complex $x-$plane and examine
the monodromy of $Q_+(x)(Q_-(x))^*$ around them.

As an example, let us seek for solutions $Q(x)$ to the Baxter
equation within the class of rational functions. The distinguished
property of such solutions is that they have a finite number of poles
on the $x-$plane and scale at infinity as an integer power of $x$.
According to \re{as} the latter property implies integer values of the
conformal weight $h$. Then, straightforward analysis shows that
$Q(x)$ can not have poles on the $x-$plane, since otherwise in order
to balance the both sides of the Baxter equation \re{Beq} the functions
$Q(x)$ should have an infinite sequence of poles shifted by $\pm i$
along imaginary axis. In similar manner, taking the limit $x\to 0$
in \re{Beq} one can argue that $Q(x)$ should vanish at the origin faster
then $x^N$. These two conditions fix $Q(x)$ to be a polynomial of
degree $h \ge N$ with $N-$times degenerate root $x=0$
\be
Q_+(x)
=x^N
\prod_{k=1}^{h-N}(x-\lambda_k)\,.
\lab{pol}
\ee
The remaining $h-N$ roots of the polynomial, $\lambda_k$, satisfy the
Bethe equations and for given $h$ they take a discrete set of real values 
\ci{FK,Bet}. Substituting this solution back into the Baxter equation \re{Beq}
and comparing its both sides one can express quantized $q_k$ in terms of the 
roots. The spectrum of $q_k$ takes the form \re{Q-gen} with $\nu_k=0$ and
$n_k=\rm odd$ and it can be calculated explicitly \ci{Bet}. Thus defined
polynomial solutions of the Baxter equation
were first obtained within the algebraic Bethe Ansatz approach and their 
properties were studied in detail \ci{Bet,Qua}.

To find the most general form of the quantization conditions \re{Q-gen}
one has to solve the Baxter equation \re{Beq} for arbitrary complex conformal
weights of the form \re{h}. Different attempts have been undertaken
\ci{MW,Qua,J}, but the analytical solution similar to \re{pol} is not 
available at present. As the first approximation to the exact solution, 
it has been proposed \ci{Int} to look for the asymptotic solution to the 
Baxter equation by applying the methods of nonlinear WKB analysis well-known 
from the studies of the Painlev\'e type-I equation \ci{P-I}. 

\subsection{WKB expansion}

Let us interpret the Baxter equation \re{Beq} as a 1-dimensional discretized
Schr\"odinger equation on the complex $x-$line and let us look for its
solutions in the form of the WKB expansion. One observes that the Planck
constant is formally unity $\hbar=1$ in the Baxter equation and the
quasiclassical limit corresponds to large values of the ``energies'' $q_k$
entering into the definition \re{Beq} of $\Lambda(x)$. This limit can be 
effectively performed by introducing an arbitrary small parameter $\eta$ and 
rescaling the coordinate, $x\to x/\eta$, and the quantum numbers, 
$q_k\to q_k\eta^k$, in \re{Beq}. Then, the asymptotic solution to the Baxter 
equation has the form \ci{PG,Qua,Int}
\be
Q(x/\eta)=\exp\lr{\frac{i}{\eta}\int^x dS(x)}\,,
\qquad
S(x)=S_0(x)+\eta S_1(x)+ \CO(\eta^2)\,,
\lab{WKB}
\ee
where $S_0(x)$ and $S_1(x)$ satisfy the relations
\be
2\cosh S_0'(x) = \frac{\widehat\Lambda(x)}{x^N}\,,\qquad
S_1'(x)=\frac{i}2 \frac{d}{dx}\ln\sinh S_0'(x)
\lab{S0}
\ee
with $S'(x)=\partial_x S(x)$ and
\be
\widehat\Lambda(x)=2x^N+\widehat q_2 x^{N-2} + ... + \widehat q_N
\,,\qquad
\widehat q_k\equiv q_k \eta^k = {\rm fixed}
\quad\mbox{as $\eta\to 0$}\,.
\lab{hat}
\ee
Similar expressions can be obtained for the antiholomorphic solution
$\widebar Q(x)$.

The limit $\eta\to 0$, or equivalently $q_k={\rm large}$, is an analog of
the classical limit in quantum mechanics. As $\eta\to 0$ the quantum 
fluctuations become frozen and the motion of particles is restricted to 
their classical trajectories defined by the leading term in \re{WKB}, the 
eikonal phase $S_0(x)$. Having identified $\sum_{k} S_0(x_k)$ as the 
``action'' function for the system of $N$ reggeons in the separated 
coordinates one may find the classical trajectories of gluons generated by 
the ``classical'' Hamiltonians $q_k$ by solving the system of the corresponding
Hamilton-Jacobi equations.  This system of equations is exactly
integrable and its solutions can be described as cycles on
the Riemann surface $\Gamma_N$ in the following way \ci{NMPZ,Int}.

Let us introduce the algebraic curve $\Gamma_N$
\be
\Gamma_N:\qquad
y^2=\widehat \Lambda^2(x)-4x^{2N}\,,
\lab{curve}
\ee
with $\widehat \Lambda(x)$ given by \re{hat}. This relation defines $y=y(x)$
as a double-valued function on the complex $x-$plane having $2N-2$
branching points at $x=\sigma_j$ such that $y(\sigma_j)=0$ or
\be
\widehat \Lambda^2(\sigma_j)=4\sigma_j^{2N}\,.
\lab{bp}
\ee
The same function becomes a single-valued on the Riemann surface
$\Gamma_N$ obtained by gluing together two sheets of the $x-$plane
along the cuts $[\sigma_{2j},\sigma_{2j+1}]$, $j=1,...,N-1$ running
between the branching points. Encircling the cuts by oriented closed
contours one constructs the set of $N-1$ cycles $\alpha_1$, $...$,
$\alpha_{N-1}$. The genus of $\Gamma_N$ equals $N-2$ and it depends on 
the number of reggeons ``inside'' the compound state.

The classical trajectories of reggeons in the separated coordinates 
$x_1$, $...$, $x_{N-1}$ correspond to the independent motion along $N-1$ 
different $\alpha-$cycles on the Riemann surface $\Gamma_N$. The branching 
points of $\Gamma_N$ become the turning points of the reggeon trajectories. 
Coming back from the separated coordinates to the original reggeon coordinates, 
$z_k$ and $\bar z_k$, one can show that the same motion looks on the 
2-dimensional plane of transverse gluon coordinates as a soliton wave 
propagating in the system of $N$ interacting gluons \ci{Int}. Its explicit 
expression was constructed in \ci{Sol} using the methods of the finite-gap 
theory out of the curve $\Gamma_N$. The quantum numbers of the $N$ reggeon 
state, $q_k$, become parameters of the soliton solutions. The quantization 
of $q_k$ appears as a result of imposing the Bohr-Sommerfeld quantization 
conditions on the periodic reggeon trajectories in the action-angle variables 
$(a_k,\varphi_k)$ defined as \ci{NMPZ}
\be
a_k=\frac1{2\pi}\oint_{\alpha_k} dS_0(x)\,,\qquad
\varphi_k=\frac{\partial S_0(x)}{\partial a_k}\,,
\lab{act}
\ee
where $0\le \varphi_k\le 2\pi$ and $\alpha_k$ are cycles on the
Riemann surface \re{curve}.

To establish the explicit form of the WKB quantization conditions let us 
consider the properties of the differential 
$dS=dS_0(x)+\eta dS_1(x)+\CO(\eta^2)$
defined by the WKB ansatz \re{WKB}. One finds from \re{S0} that $dS_0$ is the
following meromorphic differential on $\Gamma_N$
\be
dS_0(x) = dx\ln \omega(x)
\simeq \frac{dx}{y} \left[N\widehat\Lambda(x)-x\widehat\Lambda'(x)\right].
\lab{S0-eq}
\ee
Here the notation was introduced for the curve $\omega=\omega(x)$
\be
\omega+1/{\omega}=\widehat\Lambda(x)/x^N\,,\qquad
y=x^N\lr{\omega-1/{\omega}}
\lab{omega}
\ee
and the sign $\simeq$ means equivalence up to the exact differential,
$d(x\ln \omega)$, having vanishing $\alpha-$periods.

Solving \re{S0} we obtain two different solutions, $dS_{0,+}(x)$ and
$dS_{0,-}(x)$, which are continuous functions of complex $x$ except
across the cuts $[\sigma_{2j},\sigma_{2j+1}]$, satisfying the relation
$dS_{0,+}(x)=-dS_{0,-}(x)$. Being combined together they define the
differential $dS_0(x)$ as a double-valued function on the complex
$x-$plane. Nevertheless, $dS_0$ is a single-valued function
on the Riemann surface $\Gamma_N$ built from two copies of the $x-$plane
to which we will refer as upper and lower sheets. Examining \re{S0} as
$x\to\infty$ we obtain the behaviour of $dS_0(x)$ on the upper, $dS_{0,+}$,
and lower, $dS_{0,-}$, sheets of $\Gamma_N$ as
\be
dS_{0,\pm}(x) \sim \pm i\eta\sqrt{-q_2}\frac{dx}{x}
\,,
\qquad
\mbox{as $x\to\infty$}\,,
\lab{S0-inf}
\ee
where $q_2$ is given by \re{q2}.

The first nonleading correction to the WKB phase $S(x)$ is defined by the
differential $dS_1$ which one calculates using \re{S0-eq} and \re{omega} as
\be
dS_1(x)=\frac{i}4d\ln\frac{\widehat \Lambda^2(x)-4x^{2n}}{4x^{2N}}
       =\frac{i}4 dx\lr{\sum_{k=1}^{2N-2}\frac1{x-\sigma_k}-\frac{2N}{x}}
\lab{S1}
\ee
with $\sigma_k$ being the branching points \re{bp}. In contrast with $dS_0$,
this differential is well defined on the complex $x-$plane. It has poles
located at the branching points \re{bp}, at the origin and at the
infinity where its asymptotics is given by
\be
dS_1(x) \sim -\frac{i}2 \frac{dx}{x}\,,\qquad
\mbox{as $x\to\infty$}\,.
\lab{S1-inf}
\ee
Combining together \re{S0-eq} and \re{S1} we find that two different branches of
the differential $dS_0$ give rise to two linear independent WKB solutions to
the Baxter equation
$$
Q_\pm(x/\eta)\equiv\exp\lr{\frac{i}{\eta}\int^x dS_\pm(x)}\,,
$$
where
\baa
dS_+(x)&=&dS_{0,+}(x)+\eta dS_1(x)+\CO(\eta^2)\,,
\\
dS_-(x)&=&dS_{0,-}(x)+\eta dS_1(x)+\CO(\eta^2)
       =-dS_{0,+}(x)+\eta dS_1+\CO(\eta^2)\,.
\eaa
Their asymptotics at infinity can be obtained from \re{S0-inf} and
\re{S1-inf} as
$$
dS_\pm(x)=-i\eta\frac{dx}{x}\lr{\pm\sqrt{-q_2}+\frac12+\CO(\eta)}\,.
$$
We recall that at small $\eta$ the quantum numbers $q_k$ scale as
$q_k\sim 1/\eta^k$. Therefore, replacing $\sqrt{-q_2}=h-\frac12+\CO(h^{-2})$
and taking the limit $\eta\to 0$ one checks that the functions
$Q_\pm(x/\eta)$ behave at
infinity in accordance with \re{as}.

Substituting the obtained WKB expressions for $Q_\pm(x/\eta)$ into \re{ans}
and keeping only the first term in the r.h.s.\ of \re{ans} one finds
the holomorphic and antiholomorphic WKB solutions to the Baxter equation
as
\ba
Q(x/\eta)&=&\exp\left\{\frac{i}{\eta}\int^x dS_{0,+}+i\int^x dS_1+\CO(\eta)
                \right\}
\lab{Q-sol}
\\
\widebar Q(x^*/\eta)&=&\exp\left\{\frac{i}{\eta}
\lr{\int^x dS_{0,+}}^*-i\lr{\int^x dS_1}^*+\CO(\eta)\right\}
\,.
\nonumber
\ea
Let us examine the analytical properties of these functions on the
complex $x-$plane. Relations \re{S0-eq} and \re{S1} imply that the singularities
of both functions are located on the cuts $[\sigma_{2k},\sigma_{2k+1}]$. For
the product $Q(x/\eta)\widebar Q(x^*/\eta)$ to be a uniform function of $x$
one has to require that for $x$ encircling the cuts
$[\sigma_{2k},\sigma_{2k+1}]$ on the complex plane
along the cycles $\alpha_k$ it should come to the starting point with the
same value. This gives the following set of $N-1$ quantization conditions
\be
\Re\lr{\oint_{\alpha_k}dS_{0,+}}
 - \eta \Im\lr{\oint_{\alpha_k}dS_1}+ \CO(\eta^2) =\eta\pi n_k
\lab{quant}
\ee
where $n_k$ $(k=1,...,N-1)$ are integers and $\CO(\eta^2)$ stands for 
nonleading terms in the WKB expansion. According to \re{S1}, the
differential $dS_1$ has poles on the $x-$plane and the calculation of its
$\alpha-$periods is reduced to taking the residues at the
poles located inside $\alpha_k$. Since by definition each $\alpha-$cycle
encircles two branching points, $\sigma_{2k}$ and $\sigma_{2k+1}$,
and, additionally, one of the cycles, say $\alpha_{k_0}$, encircles the
pole at $x=0$, one gets
\be
\frac1{2\pi}\oint_{\alpha_k} dS_1=-\frac12+\delta_{k,k_0}\frac{N}2\,.
\lab{S1-mon}
\ee
Substituting this expression into \re{quant} we write the quantization
condition in the following form
\be
\frac1{2\pi}\oint_{\alpha_k} d S_{0,+}(x)=\eta\, \lr{\frac{n_k}2+i\nu_k}
+\CO(\eta^2)\,,
\lab{S0-mon}
\ee
with $\nu_k$ $(k=1,...,N-1)$ being arbitrary real numbers. This relation
has a simple interpretation in terms of the action variables \re{act}.
Namely, it states that the possible values of the action variables $a_k$ 
match into the spectrum of the conformal weights \re{h} of the principal 
series representation of the $SL(2,\IC)$ group.

The periods of the differential $dS_{0,+}$ entering the l.h.s.\ of the
quantization conditions \re{S0-mon} are uniquely defined by the values of the
quantum numbers $q_2$, $...$, $q_N$. Therefore, solving the system
of $N-1$ equations \re{S0-mon}, one will get $q_2$, $...$, $q_N$ as functions
of $N-1$ complex numbers $\ell_k=\frac{n_k}2+i\nu_k$, each corresponding
to $N-1$ cycles on the Riemann surface $\Gamma_N$. This result is
in agreement with our expectations \re{Q-gen} and, additionally, it implies 
that the WKB quantized $q_k$ are holomorphic functions of $\ell_k$
\be
q_k=q_k(\ell_1,...,\ell_{N-1})\,,\qquad
\ell_k=\frac{n_k}2+i\nu_k\,.
\lab{hol}
\ee
The same parameters determine the value of the conformal weight $h$ as follows.
One takes into account that the sum of the $\alpha-$periods of the differential 
$dS_{0,+}$ is equal to its period around the infinity 
where the differential has a simple pole \re{S0-inf} with the residue 
$\eta\sqrt{-q_2}$. In this way one gets using \re{S0-mon}
\be
\sqrt{-q_2}=\sum_{k=1}^{N-1} \lr{\frac{n_k}2 + i\nu_k} + \CO(\eta)
\lab{q2-sqrt}
\ee
with $\eta \sim 1/\sqrt{-q_2}$. For large $-q_2$, or equivalently large
conformal weights $h$, one replaces $\sqrt{-q_2}=h-\frac12+\CO(1/h)$
to obtain the expression for $h$ which reproduces possible quantized
values of the conformal weight \re{h} with $n=\sum_k n_k$ and
$\nu=\sum_k\nu_k$.

Having defined in \re{S1-mon} and \re{S0-mon} the periods of the
differentials, one calculates the monodromy of the WKB solutions to the
Baxter equation, \re{Q-sol}, along the $\alpha_k-$cycles as
\be
Q(x)
\stackrel{x\circlearrowleft\alpha_k}{\longrightarrow}
Q(x) \e^{2\pi i\lr{\frac{n_k-1}2+i\nu_k+\frac{N}2\delta_{k,k_0}}}\,,
\qquad
\widebar Q(x^*)\stackrel{x\circlearrowleft\alpha_k}{\longrightarrow}
\widebar Q(x^*) \e^{2\pi i\lr{\frac{n_k+1}2-i\nu_k-\frac{N}2\delta_{k,k_0}}}
\,,
\lab{Q-mon}
\ee
where higher order WKB corrections were neglected. One verifies that the
monodromies cancel against each other in the product $Q(x)\widebar Q(x^*)$.
Moreover, moving $x$ around all $\alpha-$cycles on the complex $x-$plane and
subsequently applying \re{Q-mon} one reproduces the monodromy at infinity
$$
Q(x)\to Q(x)\e^{2\pi ih}
\,,\qquad
\widebar Q(x^*)\to\widebar Q(x^*)
\e^{2\pi i (1-h^*)}\,.
$$
Relations \re{Q-mon} allow us to establish the correspondence between
WKB quantization conditions \re{S0-mon} and the special class of exact
polynomial solutions to the Baxter equation \re{pol}. Namely,
comparing \re{Q-mon} with \re{pol} one has to require that
$\exp({2\pi i\lr{\frac{n_k-1}2+i\nu_k+\frac{N}2\delta_{k,k_0}}})=1$. This
leads to the additional constraints on the parameters of the WKB
quantization conditions \re{S0-mon}:
$$
\nu_k=0\,,\qquad
n_k+\delta_{k,k_0}N=2\IZ_++1
$$
and, as a consequence, the conformal weight 
$h=\frac12+\sum_j \frac{n_j}2+i\nu_j$ takes only positive 
integer values. The resulting quantization conditions were studied in 
\ci{Qua,Int} and their solutions, $q_k$, were found to be in a good 
agreement with numerical calculations within the algebraic Bethe 
Ansatz approach \ci{Bet}.

\section{Whitham flow}

Let us apply the quantization conditions \re{S0-mon} to obtain the
expressions for the charges $q_3$, $...$, $q_N$ in the leading order
of the WKB expansion. In the leading order one neglects
$\CO(\eta^2)-$corrections to \re{S0-mon} and uses invariance of
the differential \re{S0-eq} and the curve \re{curve} under transformations
$$
x\to \lambda x\,,\qquad
\widehat q_k \to \lambda^k \widehat q_k\,,\qquad
dS_0  \to \lambda dS_0
$$
induced by reparameterization of the WKB expansion parameter,
$\eta\to\lambda\eta$, to put $\eta=1/\sqrt{-q_2}$, or equivalently
$\widehat q_2=-1$, in \re{S0-mon}. This amounts to replacing $\widehat q_k$ 
in the definition, \re{hat} and \re{S0-eq}, of the differential $dS_0$ by 
invariant parameters
\be
u_k=q_k/(-q_2)^{k/2}
\,,\qquad
k=3,...,N\,,
\lab{moduli}
\ee
called the moduli of the curve $\Gamma_N$. One also
introduces notation for the r.h.s.\ of \re{S0-mon} as
\be
\delta_k=\frac{\frac{n_k}2+i\nu_k}{\sqrt{-q_2}}\,,\qqquad
\sum_{k=1}^{N-1} \delta_k = 1\,,
\lab{delta's}
\ee
where the last relation follows from \re{q2-sqrt}.

In terms of new variables the quantization conditions \re{S0-mon}
become the relations between the moduli $u_k$ and the parameters $\delta_j$.
In particular, in the case of $N=3$ states the quantization
conditions for the moduli $u_3\equiv u$ look like
\be
\oint_{\alpha_k}\frac{(3u-2x)dx}
{\sqrt{(u-x)(4x^3-x+u)}}=2\pi\delta_k\,,\qquad k=1\,,2\,.
\lab{qc3}
\ee
Solving \re{qc3} and \re{S0-mon} one finds the moduli,
$u_k=u_k(\delta_1,...,\delta_{N-1})$, as smooth functions of the
complex parameters $\delta_k$ and then calculates the corresponding
values of the quantum numbers $q_3$, $...$, $q_N$ using \re{moduli} and
\re{delta's}.

Let us consider the relations \re{moduli} and \re{delta's} in two special
cases discussed before Sect.2.1:  $n_k=0$, $\nu_k\neq 0$ and
$n_k\neq 0$, $\nu_k=0$ for all $k$, to which we will refer below as I and II,
respectively. We find that in both cases the parameters \re{delta's}
take {\it real\/} values
$$
\delta^{\rm (I)}_k=\frac{\nu_k}{\nu_1+...+\nu_{N-1}}\,,\qqquad
\delta^{\rm (II)}_k=\frac{n_k}{n_1+...+n_{N-1}}\,.
$$
They are discrete for $\delta^{\rm (II)}_k$ and continuous for
$\delta^{\rm (I)}_k$ with the former being the subset of the latter.
Calculating the moduli as $u^{\rm (I)}_k=u_k(\delta^{\rm (I)})$
and $u^{\rm (II)}_k=u_k(\delta^{\rm (II)})$ one realizes that they enjoy
the same properties. Among other things this means that in both cases 
the moduli are universal, that is they are given by the values of
{\it same\/} function $u=u(\delta)$ on the real $\delta-$axis. In
particular, this function can be calculated for real $\delta$ in the 
case II using available exact polynomial solutions of the Baxter
equation and then applied to evaluate the quantized $q_k$ in 
the case I. 

Universality of moduli leads to the
following expressions for the quantum numbers \re{moduli} and
\re{delta's}
\be
q_2^{\rm (I)}=-(i\nu)^2\,,\qqquad
q_k^{\rm (I)}=(i\nu)^k u_k\lr{\frac{\nu_1}{\nu},...,\frac{\nu_{N-1}}{\nu}}
\lab{I}
\ee
and
\be
q_2^{\rm (II)}=-\lr{\frac{n}2}^2\,,\qqquad
q_k^{\rm (II)}= \lr{\frac{n}2}^k u_k\lr{\frac{n_1}{n},...,\frac{n_{N-1}}{n}}
\lab{II}
\ee
with $\nu=\sum_{k=1}^{N-1}\nu_k$ and $n=\sum_{k=1}^{N-1}n_k$. One notices
that the substitution
$$
\frac{n_k}2 \rightleftharpoons i\nu_k
$$
maps one of the relations into another one and the same
transformation leaves invariant the expression \re{hol} and the
quantization conditions \re{S0-mon}.

We would like to stress that the relations \re{I} and \re{II} were obtained 
in the leading order of the WKB expansion and they are valid for large $n$ 
and $\nu$. In this limit, one finds using \re{q2} that the
conformal weight has the form
\be
h^{\rm (I)}=\frac12+i\nu\,,\qqquad
h^{\rm (II)}=\frac12+\frac{n}2\,,
\lab{I+II}
\ee
which is in agreement with the general expression \re{h}.
In the case II for $n={\rm odd}$ one compares \re{II} with
the results of the polynomial solutions of the Baxter equation \ci{Qua}
to deduce that the quantized $q_k^{\rm (II)}$ take real values. As a
consequence, $u_k$ is a {\it real\/} function of its arguments.
Applying this property to \re{I} one finds that the quantum numbers
$q_k^{\rm (I)}$ are real for even $k$ and pure imaginary
for odd $k$. 
It is for these values of the quantun numbers that one expects to find 
the state in the spectrum of the effective QCD Hamiltonian \re{BKP} with 
the maximal energy \ci{WJ}.

\subsection{Whitham equations}

Let us solve the quantization conditions \re{S0-mon} and \re{qc3} and 
determine the functions $u_k$. The remarkable property of the moduli
$u_k$ is that their dependence on $\delta_j$ is governed by the Whitham
equations \ci{Wh}. To derive these equations one
varies the both sides of the quantization conditions \re{S0-mon} with 
respect to $\delta_k$ and observes the following property 
of the differential \re{S0-eq}
$$
\frac{\partial dS_0}{\partial\delta_k}
=\frac{dx}{\omega}
\frac{\partial\omega}{\partial\delta_k}
=\frac{dx}{y}
\frac{\partial\widehat\Lambda(x)}{\partial\delta_k}
=\frac{dx}{y}\sum_{j=3}^N x^{N-j}
\frac{\partial u_j}{\partial\delta_k}\,,
$$
where $\widehat\Lambda(x)=2x^N-x^{N-2}+u_3 x^{N-3}+...+u_N$ and the
relation \re{omega} was used. This leads the following system
of Whitham equations \ci{Int}
\be
\frac{\partial u_j}{\partial \delta_k}=U_{kj}(u_3,...,u_N)\,,
\lab{Wh}
\ee
with $j=3,...,N$, $k=1,...,N-1$ and $U_{kj}$ being the matrix inverse to
$$
\lr{U^{-1}}_{jk}=\frac1{2\pi}\oint_{\alpha_k}
\frac{dx\,x^{N-j}}{\sqrt{\widehat\Lambda^2(x)-4x^{2N}}}\,.
$$

The Whitham equations have the following interpretation \ci{Wh} in terms of
``classical'' reggeon trajectories defined by the action function
$S_0(x)$. The moduli $u_k$ enter as parameters into the classical soliton
wave solutions of the Hamilton-Jacobi equations. Equations \re{Wh}
describe the flow of the moduli $u_k$ in the slow ``times''
$\delta_1$, $...$, $\delta_{N-1}$, corresponding to adiabatic
deformations of the soliton waves propagating in the system of $N$
reggeons.

Analysis of the Whitham equations depends on the number of reggeons
$N$ inside the compound state. In what follows we restrict our consideration
to $N=3$ reggeon states and its generalization to higher states, $N\ge 4$,
becomes straightforward. In this case, \re{Wh} takes the form
\be
\frac{\partial\delta_k}{\partial u}=\frac1{2\pi}\oint_{\alpha_k}\frac{dx}
{\sqrt{(u-x)(4x^3-x+u)}}\,,\qquad \delta_1+\delta_2=1\,,
\lab{Wh3}
\ee
where one integrates along two $\alpha-$cycles of the elliptic curve
$\Gamma_3$ defined in \re{curve}
\be
\Gamma_3: \qquad
y^2=(u-x)(4x^3-x+u)= - 4 \prod_{j=1}^4 (x-\sigma_k)\,.
\lab{curve3}
\ee
The relations \re{Wh3} define $\delta_k$ as smooth functions on the
complex $u-$plane with the only possible singularities at the exceptional
points $u=u_{\rm crit}$, at which two branching points merge,
$\sigma_j=\sigma_k$, and the integrand in \re{Wh3} and \re{qc3}
develops a pole. It is easy to see that these points are located
on the real $u-$axis at
\be
u_{\rm crit} = -\frac1{\sqrt{27}}\,,\quad 0\,,\quad \frac1{\sqrt{27}}\,.
\lab{sing}
\ee
The integration in \re{Wh3} goes along $\alpha-$cycles that
encircle two cuts running between the branching points
$\sigma_j$ on the complex $x-$plane. Making the choice of the cuts
one requires that the moduli $u$ has to be a real function
of the flow parameter $\delta$. 

\subsection{Real moduli}

In order to calculate the quantum numbers $q_2$ and $q_3$ using
\re{I} and \re{II} it is sufficient to analyze \re{Wh3} only for real 
values of the moduli $u$ and the flow parameters $\delta_k$. Let us 
specify the location of the branching points of the curve $\Gamma_3$ 
with real $u$ on the complex $x-$plane. According to \re{curve3}, three 
branching points satisfy the equation $4\sigma_j^3-\sigma_j+u=0$ and the 
last one is $\sigma_4=u$. Introducing two parameters
\be
m=\frac{\sigma_2-\sigma_1}{\sigma_3-\sigma_1}\,,\qquad
\lambda^2=\frac19(\sigma_1-\sigma_3)^2=\frac1{12(m^2-m+1)}\,,
\lab{m}
\ee
one can parameterize the branching points in the standard way as \ci{AS}
\be
\sigma_1=-\lambda(m+1)\,,\qquad
\sigma_2=\lambda(2m-1)\,,\qquad
\sigma_3=\lambda(2-m)\,.
\lab{sigma's}
\ee
The remaining branching point, $\sigma_4=u$, gives the value of the moduli
\be
u=-4\sigma_1\sigma_2\sigma_3=\frac1{6\sqrt 3}
\frac{(2-m)(2m-1)(m+1)}{(m^2-m+1)^{3/2}}\,.
\lab{u}
\ee
One notes that the expression for $u^2$ is invariant under modular
transformations, $m\to 1/m$ and $m\to 1-m$, induced by permutations
of $\sigma_j$.

Defining the effective discriminator
$$
\Delta_{\rm eff}(u) = 16 \sigma_4^2 \prod_{i>j} (\sigma_i-\sigma_j)^2
=u^2 (1-27 u^2)\,.
$$
one has to distinguish three different cases corresponding to
$\Delta_{\rm eff}>0$, $\Delta_{\rm eff}=0$ and $\Delta_{\rm eff}<0$.
Additionally using the symmetry of $\Delta_{\rm eff}(u)$ under
$u\to -u$ one restricts the moduli to be nonnegative, $u\ge 0$.

In the first case, $\Delta_{\rm eff}>0$, or $0 < u <
\frac1{\sqrt{27}}$, one finds that all four branching points are
real and they can be ordered on the real axis as
$\sigma_1<0<u<\sigma_2<\sigma_3$. In parameterization \re{m} and
\re{sigma's}, this corresponds to $\frac12 < m < 1$.
The cuts of the function $y(x)$ defined in \re{curve3} run along
the segments $[\sigma_1,u]$ and $[\sigma_2,\sigma_3]$ on the real
axis and two $\alpha-$cycles encircle them on the complex $x-$plane
as shown in Fig.~1a.

In the second case, $\Delta_{\rm eff}=0$, two branching points merge:
either at $\sigma_2=\sigma_4=0$ for $m=1/2$, or at $\sigma_2=\sigma_3
=\frac1{2\sqrt 3}$ for $m=1$, generating two singularities on the moduli
space, $u=0$ and $u=\frac1{\sqrt{27}}$, respectively.

In the third case, $\Delta_{\rm eff}<0$, or $\frac1{\sqrt{27}}<u<\infty$,
two branching points, $\sigma_1$ and $\sigma_4=u$, are real and two
remaining ones are complex conjugated to each other, $\sigma_2^*=\sigma_3$.
This corresponds to choosing the parameters as
\be
m=\frac{\nu+i}{\nu-i}
\,,\qquad
\sigma_1=-\frac{\sqrt 3\nu}{3\sqrt{\nu^2-3}}\,,
\qquad
\sigma_2= \frac{\sqrt 3}{6}\frac{\nu+3i}{\sqrt{\nu^2-3}}\,,
\qquad
\sigma_3=\sigma_2^*
\lab{para}
\ee
with $\nu\ge \sqrt{3}$ and the expression for moduli looks like
\be
u=\frac1{\sqrt{27}}\frac{\nu(9+\nu^2)}{(\nu^2-3)^{3/2}}\,.
\lab{u-nu}
\ee
The restriction $\nu\ge \sqrt{3}$ comes from the condition for $u$ to
be real. Choosing the cuts between the branching points we require the
r.h.s.\ of \re{Wh3} and \re{qc3} to be real. As a result, one cut runs
along the segment $[\sigma_1,u]$ on the real axis and the second one
connects the points $\sigma_2$ and $\sigma_3$ on the complex plane.
The definition of the corresponding $\alpha-$cycles is shown in Fig.~1b.

\begin{figure}[ht]
\centerline{\epsffile{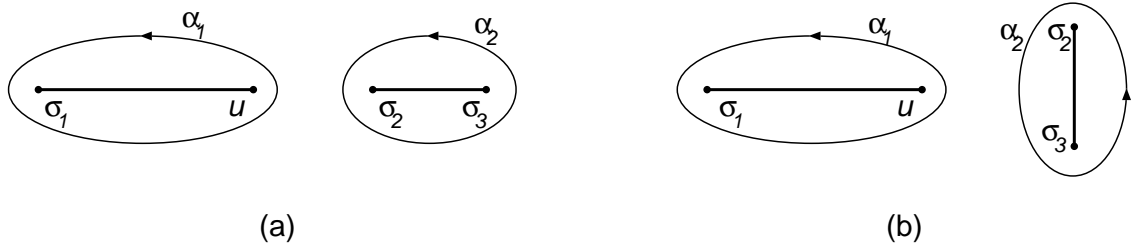}}
\caption{The definition of the $\alpha-$cycles for real positive
         moduli: (a) $0 \le u\le \frac1{\sqrt{27}}$ and (b)
         $u > \frac1{\sqrt{27}}$.}
\end{figure}

In the above consideration the moduli were assumed to be nonnegative.
To describe negative values of moduli, $u<0$, we use invariance of the
curve \re{curve3} under $x\to -x$ and $u\to -u$. This transformation
interchanges the $\alpha-$cycles on the complex $x-$plane and leads to
the relation $u(\delta_1,\delta_2)=-u(\delta_2,\delta_1)$ with
$\delta_1+\delta_2=1$. Therefore, as a function of $\delta\equiv\delta_1$
the moduli $u(\delta)\equiv u(\delta,1-\delta)$ satisfies the condition
\be
u(1-\delta)=-u(\delta)\,,
\lab{sym}
\ee
which maps positive and negative values of $u$.

\subsection{Reference points}

Before integrating the Whitham equations \re{Wh3} let us consider the
quantization conditions \re{qc3} at three reference points on the moduli
space, $u=0$, $u=\frac1{\sqrt{27}}$ and $u\to\infty$, and find the
corresponding values of the flow parameter $\delta$.

For $u=0$, or equivalently $m=\frac12$, we get from \re{qc3}
\be
\delta(0)=\frac1{\pi}\oint_{\alpha_1}\frac{dx}{\sqrt{1-4x^2}}
=\frac2{\pi}\int_{-1/2}^0\frac{dx}{\sqrt{1-4x^2}}=\frac12\,.
\lab{ref1}
\ee
The same result can be also deduced from \re{sym}. The value of moduli,
$u(\frac12)=0$, corresponds to one of the singularities \re{sing} of the
curve $\Gamma_3$, at which two cuts $[\sigma_1,u]$ and $[\sigma_2,\sigma_3]$
merge as $\sigma_2\to u$ (see Fig.1a).

For $u=\frac1{\sqrt{27}}$, or equivalently $m=1$, one approaches another
singularity \re{sing}, at which the cycle $\alpha_2$ is shrinking into
a point. Two branching points merge, $\sigma_2=\sigma_3=
\frac1{2\sqrt 3}$ and the integrand \re{qc3} develops a pole at
$x=\frac1{2\sqrt 3}$. It is compensated however by vanishing numerator
leading to
\be
\delta\lr{\mbox{$\frac1{\sqrt{27}}$}}=\frac1{2\pi}
\oint_{\alpha_1} \frac{dx}{\sqrt{\lr{x+\frac1{\sqrt 3}}
\lr{\frac1{\sqrt{27}}-x}}}
=\frac1{\pi}\int_{-\frac1{\sqrt 3}}^{\frac1{\sqrt{27}}}
 \frac{dx}{\sqrt{\lr{x+\frac1{\sqrt 3}}
 \lr{\frac1{\sqrt{27}}-x}}}
=1\,.
\lab{ref2}
\ee
Applying the symmetry \re{sym} one also finds that
$\delta\lr{-\frac1{\sqrt{27}}}=0$.

For $u\to\infty$, or equivalently $\nu\to \sqrt 3$ in \re{u-nu}, one
rescales $x\to (u/2)^{1/3} x$ in \re{qc3} and obtains the following
asymptotics
\be
\delta(u\to\infty) =
\frac3{\pi} \lr{\frac{u}2}^{1/3}
\oint_{\alpha_1} \frac{dx}{\sqrt{1+x^3}}
=\frac6{\pi}\lr{\frac{u}2}^{1/3}\int_{-1}^\infty \frac{dx}{\sqrt{1+x^3}}
=u^{1/3} \frac{2\pi}{\Gamma^3\lr{\frac23}}\,,
\lab{ref3}
\ee
which is valid up to subleading in $u$ terms.

Relations \re{ref1}, \re{ref2} and \re{ref3} determine the values of the
flow parameter corresponding to two singularities \re{sing} on the moduli
space
and fix
its asymptotics at infinity:
\be
u(\mbox{$\frac12$})=0 \,,\qquad
u(1)=\frac1{\sqrt{27}} \,,\qquad
u(\delta)\stackrel{\delta\to\infty}{=}\frac{\Gamma^9\lr{\frac23}}{(2\pi)^3}\delta^3\,.
\lab{rp}
\ee
Let us now apply the Whitham equations \re{Wh3} to reconstruct the flow
$u=u(\delta)$ between these reference points. We shall consider separately
two branches, $ 0\le u < \frac1{\sqrt{27}}$ for $\frac12\le \delta <1$
and $ u > \frac1{\sqrt{27}}$ for $\delta> 1$,
and then glue them together at $u=\frac1{\sqrt{27}}$.

\subsection{Whitham flow for $0 \le u \le \frac1{\sqrt{27}}$}

For real positive moduli inside the interval $0 \le u \le \frac1{\sqrt{27}}$
the definition of the $\alpha-$cycles is shown on Fig.~1a. The Whitham
equation \re{Wh3} looks like
\be
\frac{\partial \delta}{\partial u}=\frac1{\pi} \int_{\sigma_1}^u
\frac{dx}{\sqrt{(u-x)(4x^3-x+u)}}\,,
\lab{delta}
\ee
where $\delta\equiv\delta_1$ and the branching point $\sigma_1$ takes
real values defined in \re{sigma's}. The elliptic integral entering the
r.h.s.\ of \re{delta} is explicitly positive thus implying that $u$
is an increasing single valued function of $\delta$. It can be expressed
in terms of the elliptic function of the 1st kind defined in the
Appendix, \re{IK-def}, as  
\be
\frac{\partial \delta}{\partial u}=\frac{\IK(z)}{\pi\sqrt{c}}\,.
\lab{IK}
\ee
Here the following notations were introduced
\be
c=\frac{(2-m)^3 m}{12(m^2-m+1)^2}\,,\qquad
z=\frac{(m+1)^3(m-1)}{m(m-2)^3}\,,
\lab{c-z}
\ee
the parameter $m$ was defined in \re{m} and for the moduli inside the interval
$0 \le u \le \frac1{\sqrt{27}}$ its value can be restricted as
$\frac12\le m\le 1$.

Relation \re{IK} defines the derivative $\frac{\partial \delta}{\partial u}$
as a function of the parameter $m$. Replacing $u$ by its expression \re{u}
one integrates \re{IK} to obtain the dependence of the moduli on the
flow parameter through the parametric dependence $u=u(m)$ and $\delta=
\delta(m)$ on the interval $\frac12 \le m \le 1$. In this way,
one reconstructs the branch of the curve $u=u(\delta)$ for
$0 \le u \le \frac1{\sqrt{27}}$ and $\frac12 \le \delta \le 1$ 
shown in Fig.~2.
However, trying to obtain the explicit form of the function $u=u(\delta)$
by inverting the relations $u=u(m)$ and $\delta=
\delta(m)$ one realizes that the parameter $m$ is not a
well defined function of the moduli due to invariance of \re{u} under
modular transformations of $m$. Additionally, the end points, $m=\frac12$ and
$m=1$, correspond to the singularities on the moduli space \re{sing}, 
$u=0$ and $u=\frac1{\sqrt{27}}$, respectively, and one should expect to 
find a nonanalyticity in the dependence $u=u(\delta)$ at their vicinity.

In order to define a good parameter of the expansion of moduli which will
replace $m$ and will allow to study analyticity properties of the function $u=u(\delta)$
one introduces the dual flow parameter $\delta_D$. In analogy with
\re{qc3} it is defined as an integral of the Whitham differential $dS_0$ on
the elliptic curve $\Gamma_3$ along the $\beta-$cycle encircling 
the interval $[u,\sigma_2]$
$$
\delta_D=
\frac1{\pi}
\int_u^{\sigma_2}\frac{(3u-2x)dx}
{\sqrt{(x-u)(4x^3-x+u)}}\,.
$$
The dependence of the moduli on $\delta_D$ is
described by the Whitham equation similar to \re{delta}
\be
\frac{\partial\delta_D}{\partial u}=\frac1{\pi}
\int_u^{\sigma_2} \frac{dx}{\sqrt{(x-u)(4x^3-x+u)}}
=\frac{\IK(1-z)}{\pi\sqrt{c}}
\lab{delta-D}
\ee
with $z$ and $c$ given by \re{c-z}. The flow parameters $\delta$
and $\delta_D$ are not independent and their mutual derivative defines
the Jacobi $\tau$- and $q-$parameters of the elliptic curve $\Gamma_3$
\be
q=\e^{i\pi\tau}\,,\qqquad
\tau=i\frac{\partial\delta_D}{\partial\delta}=i\frac{\IK(1-z)}{\IK(z)}
\,.
\lab{tau}
\ee
In agreement with the Riemann theorem, $\tau$ has a positive definite
imaginary part and additionally $\Re\tau=0$ for real moduli inside the
interval $0\le u\le\frac1{\sqrt{27}}$. This property leads to
$0\le q < 1$ and it suggests us to identify the $q-$parameter 
as a good parameter of the expansion of the solutions, $u=u(q)$ and
$\delta=\delta(q)$, of the Baxter equation.

\subsubsection{Flow at the vicinity of $u=\frac1{\sqrt{27}}$}

Let us first develop the $q-$expansion of $u$ at the vicinity of
$u=\frac1{\sqrt{27}}$. This value corresponds to the parameter
$m=1$ in \re{u} and one uses \re{tau} and \re{c-z}, combined with the 
properties of the elliptic function \re{0}, to find the leading asymptotics 
of the elliptic parameters $\tau$ and $q$ as
\be
\tau=-\frac{i}{\pi}\ln\frac{1-m}2\,,
\qquad
q=\frac{1-m}2
\qquad \mbox{as $m\to 1$}\,.
\lab{q=0}
\ee
Similarly, expanding \re{u}, \re{IK} and \re{delta-D} near $m=1$ one
gets using \re{IK-0}
$$
u=\frac1{\sqrt{27}}\left[1-\frac{27}8 (1-m)^2\right]\,,\qquad
\frac{\partial\delta}{\partial m}=\frac94(1-m)\,,\qquad
\frac{\partial\delta_D}{\partial m}=-\frac{9}{4\pi}(1-m)\ln(1-m)\,,
$$
where only leading terms were kept. We conclude from these relations that
at the vicinity of $u=\frac1{\sqrt{27}}$ the moduli $u$
is an analytical function of the flow parameter $\delta$ and not
of the dual parameter $\delta_D$.

It follows from \re{q=0} that the $q-$parameter vanishes at $m=1$
and one can expand $u$ and $\delta$ into a power series in $q$. Indeed,
using \re{tau}, \re{c-z} and \re{0} to invert the dependence $q=q(m)$ 
near $m=1$
$$
1-m=2q-2q^2-2q^3+6q^4-4q^5-6q^6+\CO(q^7)
$$
and integrating \re{IK}, one obtains first few terms of the
weak coupling $q-$expansion of the moduli and the flow parameter as
\ba
u&=&\frac1{\sqrt{27}}\left[1-\frac{27}2 q^2 +\frac{891}8q^4
-\frac{12879}{16}q^6
+\frac{716499}{128}q^8
-\frac{9730773}{256}q^{10}
+\CO(q^{12})\right]
\nonumber
\\
\delta&=&1-\frac92 q^2+\frac{189}8 q^4-\frac{1917}{16} q^6
+\frac{83421}{128}q^8
-\frac{929367}{256}q^{10}
+\CO(q^{12})\,,
\lab{q-exp}
\ea
where only even powers of $q$ appear. Finally, one inverts the second
relation and expresses the moduli as a power series in the flow parameter
$\delta$
\be
u=\frac1{\sqrt{27}}\left[1-3\bar\delta+2\bar\delta^2-\frac29\bar\delta^3
+\frac{10}{81}\bar\delta^4+\frac{38}{243}\bar\delta^5+\CO(\bar\delta^6)
\right]\,,\qquad
\bar\delta\equiv 1-\delta\,.
\lab{delta-1}
\ee
This expression provides the solution to the Whitham equation \re{delta} at the
vicinity of $\delta=1$ and $u=\frac1{\sqrt{27}}$. One verifies that it
coincides with analogous expression first obtained within the class of
polynomial solutions of the Baxter equation \ci{Qua,Int}. Applying 
\re{sym} one can also find the behaviour of the moduli around $\delta=0$.

\subsubsection{Flow at the vicinity of $u=0$}

The series \re{delta-1} has a finite radius of convergence and for $\delta$
close to $\delta=\frac12$ one expects the dependence $u=u(\delta)$
to be drastically changed since the reference point $u(\frac12)=0$
is another singularity of the curve, \re{sing}.
One finds from \re{u} that $u=0$ corresponds to $m=\frac12$ and one
calculates the values of the Jacobi parameters of the curve, \re{tau},
using \re{c-z} and \re{1}, as $\tau=0$ and $q=1$. Thus, to approach the 
point $u=0$ on the moduli space using \re{q-exp} one has to develop the 
strong coupling expansion in $q$.

Let us show that at the vicinity of $u=0$ the strong
coupling expansion of moduli $u$ in $q=\exp(i\pi\tau)$ is equivalent to
the weak coupling expansion in the dual parameter $q_D$ defined as
\be
q_D=\e^{-\frac{i\pi}{3\tau}}\,,\qqquad
\ln q \ln q_D = \frac{\pi^2}3\,.
\lab{q-dual}
\ee
Repeating previous analysis we examine the behaviour of $\tau$ and $q_D$ near
$m=\frac12$. One gets using \re{1} and \re{c-z}
$$
\tau=-\frac{i\pi}3 \frac1{\ln\left[\frac23(m-\frac12)\right]}\,,
\qquad
q_D=\frac23\lr{m-\frac12} \to 0
$$
and observes that $q_D$ vanishes at $m=\frac12$ while $q=1$.
As $m\to \frac12$, the expansion of the Whitham equations \re{IK} and 
\re{delta-D} using \re{IK-1} leads to the following relations
$$
u=\frac23\lr{m-\frac12}\,,\qquad
\frac{\partial\delta}{\partial m}= -\frac2{\pi}
\ln\left[\frac23\lr{m-\frac12}\right]\,,\qquad
\frac{\partial\delta_D}{\partial m}=\frac{2}3 \left[1-\frac{14}3
\lr{m-\frac12}^2
\right]\,.
$$
They imply that as a function of the flow parameter $u=u(\delta)$ has a
singularity at $\delta=\frac12$
\be
\delta \sim \frac12-\frac3{\pi} u \ln u\,,\qquad
\mbox{as $u\to 0$}\,,
\lab{ln}
\ee
and at the same time $u$ is an analytical function of the dual flow
parameter $\delta_D$. This suggests to search for the moduli as a
function of the dual flow parameter $\delta_D$ and identify the dual
parameter $q_D$ as an appropriate parameter of the perturbative expansion
of the moduli around $u=0$.

Using \re{q-dual}, \re{tau}, \re{c-z} and \re{1} to invert the dependence
$q_D=q_D(m)$ near $m=\frac12$
$$
m-\frac12=\frac32 q_D-3q_D^3+\frac92 q_D^5-6q_D^7+\frac{21}2 q_D^9 -18 q_D^{11}
+\CO(q_D^{13})
$$
and integrating the dual Whitham equation
\re{delta-D}, one obtains the following relations
\ba
u&=&q_D-\frac{15}2 q_D^3+\frac{459}8 q_D^5-\frac{6547}{16}q_D^7
+\frac{339955}{128}q_D^9 +\CO(q_D^{11})
\nonumber
\\
\delta_D&=&q_D-\frac{11}2 q_D^3 +\frac{243}8 q_D^5-\frac{2599}{16}q_D^7
+\frac{113699}{128}q_D^9
+\CO(q_D^{11})\,,
\lab{qD-exp}
\ea
which provide a weak coupling expansion of the moduli and the dual flow
parameter in $q_D$. Inverting the dependence $\delta_D=\delta_D(q_D)$ we 
obtain from \re{qD-exp} the expression for $u=u(\delta_D)$ as a power 
series in the dual flow parameter $\delta_D$
$$
u=\delta_{{D}}-2\,\delta_{{D}}^{3}-6\,\delta_{{D}}^{5}-48\,
\delta_{{D}}^{7}-510\,\delta_{{D}}^{9}+\CO\left(\delta_{{D}}^{11}
\right)\,,
$$
which should be compared with \re{delta-1}. Finally, to restore the
dependence of the moduli on $\delta$, one rewrites the definitions
\re{tau} and \re{q-dual} in the form
$$
\delta=\frac12-\frac{3}{\pi} \int_0^{q_D} d q_D \ln q_D
\frac{\partial\delta_D}{\partial q_D}\,.
$$
and replaces $\delta_D$ by its expression \re{qD-exp}. This gives
the $q_D-$expansion of $\delta$ which can be combined with the
first relation in \re{qD-exp} to determine the nonleading terms
in the asymptotic expansion \re{ln}
\be
\delta\stackrel{u\to 0}{=}\frac12-\frac3{\pi}\lr{u+2u^3+18u^5}\ln u
              +\frac1{\pi}\lr{3u-\frac{11}2 u^3-\frac{1323}{20} u^5}
      +\CO(u^7) \,.
\lab{delta-1/2}
\ee
This relation can be inverted and $u$ can be expressed in terms of
the Lambert function of $\delta$.

\subsection{Whitham flow for $u > \frac1{\sqrt{27}}$}

The Whitham equation for $u > \frac1{\sqrt{27}}$ has the same form as
\re{delta}, but important difference with the previous case is that two
branching points, $\sigma_2$ and $\sigma_3$, take complex values. According
to our choice of the cuts between the branching points, shown in Fig.~1b,
the elliptic integral entering the r.h.s.\ of \re{delta} takes real
positive values defining $u$ to be an increasing single valued function
of $\delta$ for $\delta > 1$. Since $u=u(\delta)$ has the same property
for $0\le \delta < 1$ it can be now extended using \re{sym} to arbitrary
real $\delta$.

Using parameterization of the branching points \re{para} the elliptic
integral in \re{delta} can be calculated as
\be
\frac{\partial\delta}{\partial u} =\frac{2\sqrt 3}{\pi}
\frac{\nu^2-3}{(9+\nu^2)^{3/4}(1+\nu^2)^{1/4}}
\IK\lr{\frac12-\frac12\Re\left[
\frac{(\nu+i)(\nu-3i)^3}{(\nu-i)(\nu+3i)^3}
\right]^{1/2}}\,,
\lab{Wh-nu}
\ee
where $\nu\ge \sqrt 3$. Replacing $u$ by its expression \re{u-nu}  
one integrates numerically \re{Wh-nu} to restore the $\nu-$dependence 
of $\delta$ and then calculates the moduli through the
parametric dependence $u=u(\nu)$. This gives the second branch
of the curve $u=u(\delta)$ for $\delta > 1$ shown in Fig.~2.

To specify the appropriate boundary conditions for \re{Wh-nu}
one considers two values of the parameter: $\nu=\sqrt 3$ and
$\nu\to\infty$. For $\nu\to\sqrt 3$ one finds from
\re{u-nu} that $u$ infinitely increases and its leading asymptotic
behaviour in $\delta$ is given by \re{rp}. For $\nu\to\infty$ one
gets from \re{u-nu} and \re{rp} the corresponding value of moduli
as $u=\frac1{\sqrt{27}}$ for $\delta=1$. Let us consider the flow
of $u$ in both cases in more detail.

For large $\nu$ one uses \re{u-nu} and \re{Wh-nu} to expand $u$ and
$\delta$ in inverse powers of $1/\nu$
\baa
u&=&\frac1{\sqrt{27}}
\left (1+{\frac {27}{2}}\,{\nu}^{-2}+{
\frac {459}{8}}\,{\nu}^{-4}+{\frac {3375}{16}}\,{\nu}^{-6}+{\frac {
93555}{128}}\,{\nu}^{-8}+{\frac {627669}{256}}\,{\nu}^{-10}
+ \CO({\nu}^{-12})\right )
\\
\delta&=&
1+\frac92\,{\nu}^{-2}+{\frac {45}{8}}\,{\nu}^{-4}+{\frac {477}{16}}\,{\nu}
^{-6}+{\frac {189}{128}}\,{\nu}^{-8}+{\frac {110295}{256}}\,{\nu}^{-10
}+\CO({\nu}^{-12})\,.
\eaa
Inverting these relations one obtains the expansion of $u$ in powers
of $\bar\delta=1-\delta$, which identically coincides with \re{delta-1}.
This means that two different branches of the function $u(\delta)$
corresponding to the Whitham flow for $\frac12 \le \delta < 1$ and
$\delta>1$ can be smoothly glued at $\delta=1$ (see Fig.~2).

For $\nu\to\sqrt 3$ the moduli \re{u-nu} goes to infinity as
\be
u=4(\nu^2-3)^{-3/2}+(\nu^2-3)^{-1/2}+\CO\lr{(\nu^2-3)^{3/2}}\,.
\lab{u-infty}
\ee
To find the corresponding behaviour of the flow parameter one takes into
account that $\delta=1$ for $\nu=\infty$ and integrates the Whitham
equation \re{Wh-nu} as
$$
\delta=1+\int_\infty^\nu d\nu\, \frac{\partial u}{\partial\nu}\,
\frac{\partial\delta}{\partial u}\,.
$$
Replacing the derivative $\frac{\partial\delta}{\partial u}$ by its
expression \re{Wh-nu}, we expand the integral around $\nu^2=3$ and 
apply \re{number} to get after some calculations
\be
\delta=
\frac{2^{5/3}\pi}{\Gamma^3\lr{\frac23}}
\ (\nu^2-3)^{-1/2}-1+\CO\lr{(\nu^2-3)^{1/2}}\,.
\lab{delta-inf}
\ee
Combining together \re{u-infty} and \re{delta-inf} we obtain the asymptotic
behaviour of the moduli at large {\it positive\/} values of the flow
parameter $\delta$
\be
u=\frac{\Gamma^9\lr{\frac23}}{(2\pi)^3}\,
\lr{\delta^3+3\delta^2}\times
\left[1+\CO(1/\delta)\right]\,,\qquad
\mbox{as $\delta\to \infty$}\,.
\lab{del+}
\ee
Here, the leading term coincides with \re{rp}. The symmetry \re{sym} allows
to extend the flow to large negative $u$ as
\be
u=\frac{\Gamma^9\lr{\frac23}}{(2\pi)^3}\,
\lr{\delta^3-6\delta^2}\times
\left[1+\CO(1/\delta)\right]\,,\qquad
\mbox{as $\delta\to -\infty$}\,.
\lab{del-}
\ee

\begin{figure}[ht]
\vspace*{-20mm}
\centerline{\epsffile{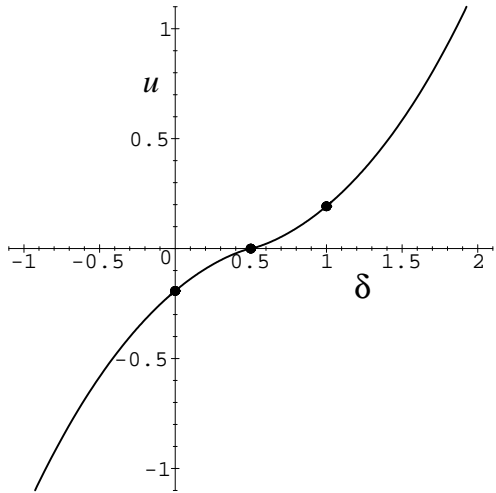}}
\vspace*{-20mm}
\caption{Quantized moduli $u$ for real values of the flow parameter
$\delta$. The dots indicate the positions of three singularities of the
elliptic curve $\Gamma_3$ located at $\delta-\frac12=\pm \frac12$ and $0$.
The behaviour of the function $u=u(\delta)$ around these points
and at infinity is described by \re{delta-1/2}, \re{delta-1} and \re{del+}.
}
\end{figure}

Summarizing the results of this Section we present on Fig.~2 the dependence 
of the quantized moduli $u=u(\delta)$ on the flow parameter 
$-\infty<\delta<\infty$. According to \re{sym} the moduli $u$ is an odd 
function of $\delta-\frac12$. Its asymptotics around $\delta=1/2$, $1$ 
and $\infty$ is described by \re{delta-1/2}, \re{delta-1} and \re{del+}, 
respectively. We recall that this curve provides the solution to 
the quantization conditions \re{qc3} and \re{Wh3} for $N=3$ reggeon 
compound states. 

\section{Quantum numbers of $N=3$ reggeon states}

Let us apply the results of the previous section to evaluate the quantum
numbers $q_2$ and $q_3$ of the $N=3$ reggeon states in the leading order of 
the WKB expansion. According to \re{hol}, the values of $q_3$ depend on 
two complex numbers $\ell_1=\frac{n_1}2+i\nu_1$ and $\ell_2=\frac{n_2}2+i\nu_2$. 
Following \re{I} and \re{II} we consider two special cases, $n_1=n_2=0$ and
$\nu_1=\nu_2=0$, corresponding to real values of the moduli and the flow
parameters.%
\footnote{To obtain real values of $u$ and $\delta$ for $N=3$ states it
          is enough to impose even weaker condition: $n_1/n_2=\nu_1/\nu_2$.}
In the case I, the relations \re{I} and \re{I+II} between the moduli
and quantized $q_3$ can be written in two equivalent forms
\be
q_3=(i\nu)^3 \times u\lr{\frac{\nu_1}{\nu}}\,,\qqquad
h-\frac12=i\nu\,,\qqquad \nu=\nu_1+\nu_2
\lab{QQ3}
\ee
or
\be
q_3=(i\nu_1)^3 \times u(\delta)\,\delta^{-3}\,,\qqquad
h-\frac12=(i\nu_1)\times \delta^{-1}\,,\qqquad
\delta=\frac{\nu_1}{\nu_1+\nu_2}\,,
\lab{Q3}
\ee
with $\nu_{1,2}$ being real. In the case II, one gets similar relations
by replacing $i\nu_k\to\frac12n_k$ and taking $n_{1,2}$ to be integer.

It is easy to see from \re{sym} that $q_3$ is antisymmetric under 
interchanging of $\nu_1$ and $\nu_2$, while the conformal weight 
$h=\frac12+i(\nu_1+\nu_2)$ is explicitly symmetric. 
Therefore, $q_3$ vanishes for $\nu_1=\nu_2$ and the
corresponding $N=3$ reggeon state with the conformal weight $h=\frac12+2i\nu_1$
is degenerate \ci{Bet}. Their wave function does not depend on one of the 
reggeon coordinates and the corresponding energy is equal to the energy of 
$N=2$ reggeon state, $E_3(q_3=0,h) = E_2(h)$.

Since quantized $q_3$ and $h$ depend on the same parameters $\nu_1$ and
$\nu_2$ one can express $\nu_2$ in terms of $h$ and consider $q_3$ to be
a function of the conformal weight $h$ and real $\nu_1$. Let us consider
separately the dependence of $q_3$ on its arguments, $q_3=q_3(h,\nu_1)$.

For fixed values of the conformal weight $h={\rm fixed}$ it is convenient
to apply \re{QQ3}. One finds that up to rescaling of the argument
the $\nu_1-$dependence of $q_3$ is governed by the moduli and it follows
the pattern of the curve shown in Fig.~2.

For fixed values of $\nu_1$ one uses instead \re{Q3}. Introducing
notations for $q=u(\delta)\,\delta^{-3}$ and $x=\delta^{-1}$
one writes \re{Q3} as
\be
q_3=(i\nu_1)^3\times q(x)\,,\qqquad
h-\frac12=i\nu_1\times x \,,\qqquad
q=x^3 u(1/x) \,,
\lab{q(x)}
\ee
with $q$ depending on $x$ and not on $\nu_1$. Here, the parameter $\nu_1$
fixes the scale of $q_3$ and $h-\frac12$. It takes real continuous values 
which cannot be however arbitrary small for the leading order WKB 
approximation to be applicable. In the case II one replaces 
$i\nu_1\to \frac12 n_1$ and uses integer $n_1$ to label the curves 
$q_3=q_3(h;n_1)$. Again, the values of $n_1$ cannot be small. 

The function $q=q(x)$ can be determined out of the curve $u=u(\delta)$ 
depicted on Fig.~2 in two steps. One first finds the dependence of moduli 
$u$ on $x=1/\delta$ as shown in Fig.~3 and then obtains $q=q(x)$ in the form 
shown in Fig.~4.

\begin{figure}[ht]
\vspace*{-20mm}
\centerline{\epsffile{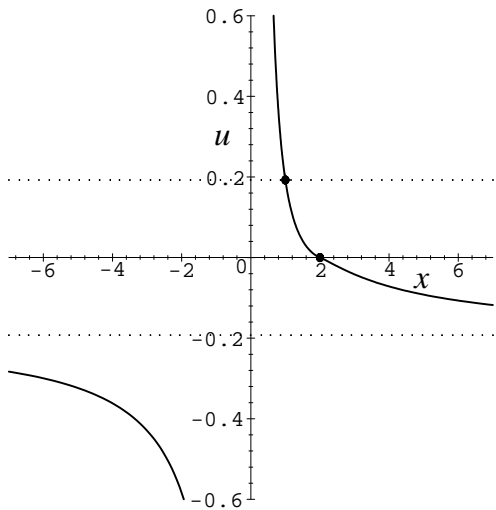}}
\vspace*{-20mm}
\caption{Two different branches of the function $u(1/x)$ entering into \re{q(x)}.
Dotted lines represent $u=\pm\frac1{\sqrt{27}}$ 
and two dots indicate the singular points $x=1$, $u=\frac1{\sqrt{27}}$ 
and $x=2$, $u=0$. The part of the curve $u^2(x)\le \frac1{27}$ corresponds 
to the polynomial solutions of the Baxter equation and it should be compared 
with Fig.~4 in \ci{Qua}.}
\end{figure}

The following comments are in order. To describe the transition from
Fig.~2 to Fig.~3 one has to take into account that $\delta=0$
is mapped into infinities $x=\pm \infty$ and two parts of the
continuous function $u=u(\delta)$ for $\delta < 0$ and $\delta > 0$ give
rise to two branches of the functions $u$ and $q$ on Figs.~3 and 4,
respectively.

\begin{figure}[ht]
\vspace*{-20mm}
\centerline{\epsffile{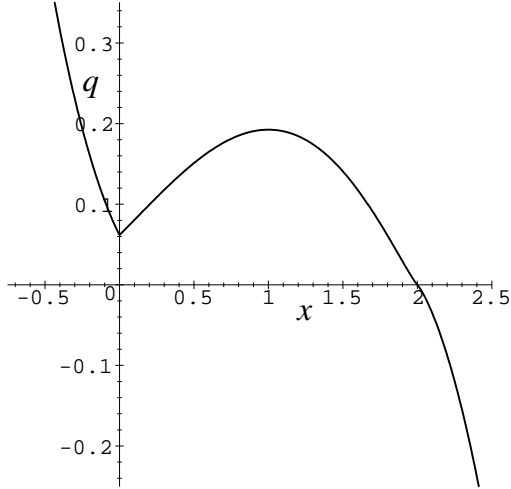}}
\vspace*{-20mm}
\caption{The function $q=q(x)$ defining the quantum numbers \re{q(x)} of 
the $N=3$ reggeon compound states. Its behaviour around $x=0_\pm$, $x=1$,
$x=2$ and $x=\pm\infty$ is given by \re{F1}, \re{F2}, \re{F4}, \re{F3} and
\re{h-infty}, respectively.}
\end{figure}

The asymptotics of $q$ as $x\to \pm\infty$ follows from \re{delta-1} and
\re{sym}
\be
q=-\frac{x^3}{\sqrt{27}} \left[1-\frac3{x}
+\CO\lr{\frac1{x^2}}
\right]\,,\quad
\mbox{as $x \to \pm\infty$}\,.
\lab{h-infty}
\ee
Two infinities $\delta \to \pm\infty$ are mapped into the origin
$x=0$. Since $u=u(\delta)$ has different subleading terms in
the asymptotics \re{del+} and \re{del-}, the function $q$
has different asymptotics as $x$ approaches the origin
from different sides (see Fig.~4)
\be
q=\frac{\Gamma^9(2/3)}{(2\pi)^3}\left[1+3x +
\CO(x^2)\right]\,,\qquad
\mbox{as $x \to 0_+$}
\lab{F1}
\ee
and
\be
q=\frac{\Gamma^9(2/3)}{(2\pi)^3}\left[1-6x +
\CO(x^2)\right]\,,\qquad
\mbox{as $x \to 0_-$}
\lab{F2}
\ee
It is this property that is responsible for the appearance of a cusp on
Fig.~4 at $x=0$ and $q=\frac{\Gamma^9(2/3)}{(2\pi)^3}=0.0617...$
Two reference points \re{ref1} and \re{ref2} correspond to
$x=2$ and $x=1$, respectively, and the asymptotics of $q$ at their
vicinity can be found from \re{ln} and \re{delta-1} as 
\be
q\ln q=\frac{2\pi}3 \lr{x - 2}\,,
\quad
\mbox{as $x\to 2$}
\lab{F3}
\ee
and
\be
q=\frac1{\sqrt{27}}\left[1-(x-1)^2
+\CO\lr{(x-1)^3}\right]\,,
\quad
\mbox{as $x\to 1$}\,.
\lab{F4}
\ee
The part of the function $u=u(\delta)$ corresponding to $0 \le \delta \le 1$ 
and $u^2 \le \frac1{27}$ describes the polynomial solutions of the Baxter 
equation. It is mapped into the branch that starts at $q=\frac1{\sqrt{27}}$ 
for $x=1$ on Fig.~4 and then decreases to infinity according to \re{h-infty} 
as $x$ increases. One checks that this behaviour is an agreement with the 
numerical results (see Fig.~4 in \ci{Qua}).

We would like to stress that the above results were obtained in the leading 
order of the WKB expansion and they are valid for large values of quantum
numbers. Examining Fig.~4 and using \re{q(x)} one notices that the latter 
condition is violated at $x=0$ and $x=2$ when either $|h-\frac12|$ or $q_3$
vanish. Therefore, one should expect that the behaviour of the function 
$q=q(x)$ close to the cusp on Fig.~4 and around the point $q=0$ will be 
modified by the nonleading WKB corrections. Their detailed analysis deserves 
further investigation.

\begin{figure}[ht]
\vspace*{-20mm}
\centerline{\epsffile{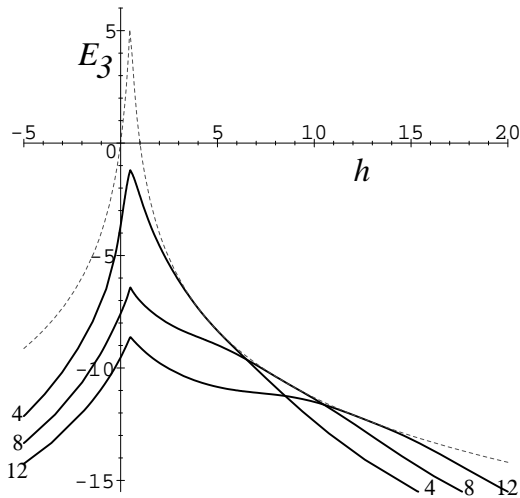}}
\vspace*{-20mm}
\caption{The energy spectrum of the $N=3$ reggeon compound states
evaluated in the leading order of the WKB quantization for type-II
quantization conditions, $E_3=E_3(h;n_1)$. Three 
different curves correspond to $n_1=4$, $8$ and $12$. 
The dotted line represents the BFKL energy of the $N=2$ reggeon state,
$E_2(h)=4\left[{\Psi(1)-\Psi(\frac12+|h-\frac12|)}\right]$.}
\end{figure}

Concluding our consideration we would like to apply the obtained results
to calculate the energy of the $N=3-$reggeon states. The corresponding
expression looks like
\be
E_3^{\rm WKB}(q_2,q_3)=-2\ln2-\sum_{j=1}^3
\left[\Psi(1+i\delta_j)+\Psi(1-i\delta_j)-2\Psi(1)\right]
\lab{E3}
\ee
where $\Psi(x)=\frac{d}{dx}\ln\Gamma(x)$ and
$\delta_j$ are defined as zeros of $\Lambda(x)$
$$
\Lambda(\delta_j)=2\delta_j^3+q_2 \delta_j+q_3=0\,.
$$
One should warn however that, first, this expression was obtained for the type-II 
solutions \re{II} and, second, it is not exact and is valid in the leading 
order of the WKB expansion. Replacing $i\nu_1\to n_1/2$ in \re{q(x)} and 
substituting quantized $q_2$ and $h$ into \re{E3} one gets $E_3=E_3(h;n_1)$ 
as a function of the conformal weight $h$ and integer $n_1$. To get an 
insight into the spectrum of $E_3$ we plot on Fig.~5 the dependence of $E_3$ 
on the conformal weight $h$ for different values of the integer $n_1$. 
These curves are in an agreement with the numerical results of the polynomial 
solutions (see Fig.~8 in \ci{Qua}) and they support the calculation of the
odderon intercept performed in \ci{Qua}. For given $n_1$ the energy is maximal
for $h=\frac12$ and its value increases as $n_1$ decreases. The absolute 
maximum of the energy $E_0={\rm max}_{h,n_1}E_3(h;n_1)$ defines the odderon 
intercept and it is expecting to occur at $n_1=1$ and $h=\frac12$. One also 
observes that for the upper curve on Fig.~5 the maximal energy, 
$E_3(\frac12;4) < E_0$, is close to the lower bound on the odderon energy 
obtained in the variational approach \ci{B}. We refer to \ci{Qua} for detailed 
description of the properties of curves shown on Fig.~5. It would be 
interesting to improve \re{E3} by applying the expression for the energy 
$E_3$ proposed in \ci{WJ}.

\section{Summary}

In this paper, we have studied the spectrum of the reggeon compound states
in high-energy QCD. These states appear as solutions of the $(2+1)-$dimensional
Schr\"odinger equation \re{BKP} which exhibits remarkable properties of the 
$SL(2,\IC)$ invariance and complete integrability. As a result, the energy of 
the states can be evaluated as a function of the set of quantum numbers $q_2$, 
$...$, $q_N$. The latter are defined as eigenvalues of the mutually commuting 
integrals of motions and their possible values are constrained by the 
quantization conditions. 

We have established the quantization conditions by applying the methods
of nonlinear WKB analysis to the $N-$reggeon Schr\"odinger equation in the
separated coordinates. In the leading order of the WKB expansion, the $N$ 
reggeon state looks like the classical system of $N$ particles moving on the 
$2-$dimensional plane of transverse gluon coordinates along periodic 
trajectories. This collective motion can be also considered as a propagation 
of the soliton wave in the system of $N$ reggeons. The charges $q_2$, $...$, 
$q_N$ become parameters of the soliton waves. 

The quantization conditions follow from the analysis of the wave function of 
the $N$ reggeon state in the leading order of the WKB expansion. Requiring
the wave function to be a single-valued function of the separated
coordinates, we have found that the selection rules for $q_2$, $...$, $q_N$ 
have the form of the Bohr-Sommerfeld quantization conditions imposed on the 
reggeon classical trajectories in the action-angle variables. The same 
conditions can be interpreted as Whitham equations on the moduli of the 
spectral curve $\Gamma_N$ corresponding to the classical $N$ reggeon system. 
We have solved the Whitham equation for $N=3$ reggeon states by using
the properties of the elliptic integrals on the curve $\Gamma_3$. 
The quantized $q_2$ and $q_3$ were obtained in the form of perturbative 
expansion in powers of the Jacobi $q-$parameter of the elliptic curve 
$\Gamma_3$. Different parts of the spectrum of $q_2$ and $q_3$ correspond 
to the weak, $q\ll 1$, and strong, $q\sim 1$, coupling regime of the 
perturbative $q-$series. In the latter case, one performs the duality 
transformation to express $q_2$ and $q_3$ as a weak coupling expansion
in the dual coupling constant, $q_D \ll 1$. Combining together different
branches we have obtained the spectrum of quantized $q_2$ and $q_3$ which 
is in agreement with available exact solutions found within algebraic
Bethe Ansatz approach.

In conclusion, one should mention that the above consideration was restricted 
to the leading order WKB expansion and one should additionally study the
importance of nonleading corrections. In particular, considering the behaviour
of \re{WKB} around the origin $x=0$ one can argue \ci{Qua} that nonleading 
WKB corrections to $Q(x)$ become equally important. As in the case of the
polynomial solutions, \re{pol}, the analysis of the singularities of the 
solutions to the Baxter equation at $x=0$ should lead to additional constaints
on the quantum numbers of the reggeon states and presumably fix the ratio 
$C_{+-}/C_{-+}$ in \re{ans}. These questions deserve additional studies.

\section*{Acknowledgements}

I would like to thank I.V. Komarov for stimulating discussions.

\setcounter{equation} 0
\def\theequation{A.\arabic{equation}}

\section*{Appendix: Elliptic integral of the first kind}

In this appendix we collect some useful properties of the elliptic
integral of the first kind. It is defined as \ci{AS,spec}
\be
\IK(z)=\int_0^1\frac{dx}{\sqrt{(1-x^2)(1-zx^2)}}
=\frac{\pi}{2} F\lr{\mbox{$\frac12$,$\frac12$};1;z}
\lab{IK-def}
\ee
with $F$ being a hypergeometric function. For $z\to 0$ it behaves as
\be
\IK(z)=\frac{\pi}2
\left\{1+\lr{\frac12}^2 z
+\lr{\frac{1\cdot 3}{2\cdot 4} }^2 z^2 + ...\right\}
\lab{IK-0}
\ee
and for $z\to 1$ as
\be
\IK(z) = -\frac1{\pi} \IK(1-z)\,\ln\frac{1-z}{16}
- 2\mbox{$\lr{\frac12}^2\frac1{1\cdot 2}$} (1-z)
- 2\mbox{$\lr{\frac{1\cdot 3}{2\cdot 4}}^2 \lr{\frac1{1\cdot 2}+\frac1{3\cdot
4}}$}(1-z)^2 + ...
\lab{IK-1}
\ee
Calculating \re{delta-inf} one uses the relation
\be
\IK\lr{\frac12+\frac{\sqrt 3}4}=\frac{3^{-3/4}2^{2/3}\pi^2}{\Gamma^3(\frac23)}
\,.
\lab{number}
\ee
The Jacobi $\tau$ and $q$ parameters defined in \re{tau}
have the following asymptotics for $z\to 0$
\ba
\tau(z) &=& \frac{i}{\pi} \ln\frac{16}{z} -\frac{i}{2\pi}
\lr{z+{\frac {13}{32}}{z}^{2}+{\frac {23}{96}}{z}^{3}+{\frac {2701}{16384
}}{z}^{4}+ \CO(z^5)}
\nonumber
\\
q(z)    &=& \frac{x}{16}\lr{1+\frac12 z+ \frac{21}{64} z^2 +\frac{31}{128} z^3
                            +\frac{6257}{32768} z^4 +\CO(z^5)}\,.
\lab{0}
\ea
and their asymptotics for $z\to 1$ can be found using the relations
\be
\tau(z)=-\frac1{\tau(1-z)}\,,\qqquad
q(z)=\exp\lr{\frac{\pi^2}{\ln q(1-z)}}\,,
\lab{1}
\ee
which follow from the definition \re{tau}.
\bb{99}
\bi{BKP}  J. Bartels, Nucl. Phys. B175 (1980) 365;
\\        J. Kwiecinski and M. Praszalowicz, Phys. Lett. B94 (1980) 413.
\bi{Lip1} L.N. Lipatov, {\it Pomeron in quantum chromodynamic\/},
          in ``Perturbative QCD'', pp.411--489, ed. A.H. Mueller,
          World Scientific, Singapore, 1989.
\bi{group}I.M. Gelfand, M.I. Graev and N.Ya. Vilenkin, {\it Generalized
          functions\/}, Vol. 5, Academic Press, 1966;
\\        D.P. Zhelobenko and  A.I. Shtern, {\it Representations of Lie
          groups\/} (in Russian), Nauka, Moscow, 1983, pp.211-220.
\bi{Lip2} L.N. Lipatov, Phys. Lett. B251 (1990) 284;  B309 (1993) 394.
\bi{FK}   L.D. Faddeev and G.P. Korchemsky, 
          [hep-ph/9404173]; Phys. Lett. B 342 (1995) 311.
\bi{Lip}  L.N. Lipatov, JETP Lett. 59 (1994) 596. 
\bi{SoV}  E.K. Sklyanin, {\it The quantum Toda chain\/},
          Lecture Notes in Physics, vol.\ 226, Springer, 1985, pp.196--233;
          {\it Functional Bethe ansatz\/}, in ``Integrable
          and superintegrable systems'', ed.\ B.A. Kupershmidt, World
          Scientific, 1990, pp.8--33;
          Progr. Theor. Phys. Suppl. 118 (1995) 35 [solv-int/9504001].
\bi{KK}   I.V. Komarov and V.V. Zalipaev, J. Phys. A: Math. Gen. 17 (1984) 1479;
\\        I.V. Komarov and V.B. Kuznetsov, J. Phys. A: Math. Gen. 23 (1990) 841.
\bi{Bet}  G.P. Korchemsky, Nucl. Phys. B443 (1995) 255.	
\bi{Qua}  G.P. Korchemsky, Nucl. Phys. B462 (1996) 333.
\bi{WJ}   J. Wosiek and R.A. Janik, Phys. Rev. Lett. 79 (1997) 2935;
          in Proceedings of the ICHEP 96, pp.615-618 [hep-th/9611025].
\bi{Tri}  G.P. Korchemsky, preprint LPTHE-Orsay-97-62 [hep-ph/9711277].
\bi{SL2}  J. Teschner, preprints LPM-97 [hep-th/9712258]; [hep-th/9712256].
\bi{Int}  G.P. Korchemsky, Nucl. Phys. B498 (1997) 68;
          in Proceedings of the ICHEP 96, pp.713-716 [hep-ph/9610454].	  
\bi{MW}   Z. Maassarani and S. Wallon, J. Phys. A: Math. Gen. 28 (1995) 6423.
\bi{J}    R. A. Janik, Acta Phys. Polon. B27 (1996) 1275.
\bi{P-I}  S.P. Novikov, Func. Anal. Appl. 24 (1990) 296;
\\        I.M. Krichever, ETH preprint, Z\"urich, June 1990;
\\        G. Moore, Comm. Math. Phys. 133 (1990) 261;
\\        F. Fucito, A. Gamba, M. Martinelli and O. Ragnisco,
          Int. J. Mod. Phys. B6 (1992) 2123.
\bi{PG}   V. Pasquier and M. Gaudin, J. Phys. A25 (1992) 5243.
\bi{NMPZ} S.P. Novikov, S.V. Manakov, L.P. Pitaevskii and V.E. Zakharov,
          {\it Theory of Solitons: The Inverse Scattering Method\/},
          Consultants Bureau, New York, 1984;
\\        B. Dubrovin, I. Krichever and S. Novikov,
          {\it Integrable systems - I},
          Sovremennye problemy matematiki (VINITI), Dynamical systems - 4
          (1985) 179;
\\        B.A. Dubrovin, V.B. Matveev and S.P. Novikov,
          Russ. Math. Surv. 31 (1976) 59.
\bi{Sol}  G.P. Korchemsky and I.M. Krichever, Nucl. Phys. B505 (1997) 387.
\bi{Wh}   G.B. Whitham, {\it Linear and Nonlinear Waves\/},
          John Wiley, New York, 1974;
\\        H. Flaschka, M.G. Forest and D.W. McLaughlin, Comm. Pure Appl.
          Math. 33 (1980) 739;
\\        S.Yu. Dobrokhotov and V.P. Maslov, J. Sov. Math. 16 (1981) 1433;
\\        B.A. Dubrovin and S.P. Novikov, Russ. Math. Surv. 44 (1989) 35.
\\        I.M. Krichever, Comm. Math. Phys. 143 (1992) 415;
          Comm. Pure Appl. Math. 47 (1994) 437;
\\        B.A. Dubrovin, Comm. Math. Phys. 145 (1992) 195.
\bi{AS}   {\it Handbook of Mathematical Functions\/}, eds. H. Abramowitz and I. Stegun, 
          Dover, New York, 1972.  
\bi{spec} {\it Higher transcendental functions\/},
          ed. A. Erd\'elyi, McGraw-Hill, 1953.
\bi{B}    M.A. Braun, preprint SPbU-IP-1998/3 [hep-ph/9801352].	  
\eb
\end{document}